\documentclass[aps,prb,twocolumn,superscriptaddress]{revtex4-1}
\usepackage{times}
\usepackage{graphicx}
\usepackage{dcolumn}
\usepackage{bm}
\usepackage{color}
\usepackage{amsmath}
\usepackage{amssymb}
\usepackage{braket}
\newsavebox{\mybox}

\usepackage[colorlinks=true]{hyperref}

\begin{document}

\title{Electronic spectral function in fractionalized Pair Density Wave scenario}

\author{M. Grandadam}

\affiliation{Institut de Physique Th\'eorique, Universit\'e Paris-Saclay, CEA, CNRS, F-91191 Gif-sur-Yvette, France}

\author{D. Chakraborty}

\affiliation{Institut de Physique Th\'eorique, Universit\'e Paris-Saclay, CEA, CNRS, F-91191 Gif-sur-Yvette, France}
\affiliation{Department of Physics and Astronomy, Uppsala University, Box 516, S-751 20 Uppsala, Sweden}\thanks{Present address}

\author{X. Montiel}

\affiliation{Department of Physics, Royal Holloway, University of London, Egham Surrey, United Kingdom}

\author{C. P\'epin}

\affiliation{Institut de Physique Th\'eorique, Universit\'e Paris-Saclay, CEA, CNRS, F-91191 Gif-sur-Yvette, France}

\begin{abstract}
Studies of the electronic spectral function in cuprates by Angle-Resolved Photo-Emission Spectroscopy reveal unusual features in the pseudogap phase that persist in the superconducting phase. We address here these observations based on the recently proposed idea that the pseudogap is due to the fractionalization of modulated particle-particle pairs (a Pair Density Wave) into uniform particle-particle and modulated particle-hole pairs. The constraint that appears between these two types of pairs can be seen has an amplitude for the pseudogap energy scale. This constraint directly modify the electronic spectral function in the pseudogap phase. We derive a self-consistent equation for the pseudogap amplitude and show that it leads to the formation of Fermi arcs. The band dispersion obtained in the anti-nodal region is in good agreement with experimental ARPES observations in Pb$_{0.55}$Bi$_{1.5}$Sr$_{1.6}$La$_{0.4}$CuO$_{6+\delta}$ (Bi2201) and present a back-bending that goes to the Fermi level as we go away from the antinodal region. We also discuss the temperature dependence of the ARPES spectrum in the pseudogap and in the superconducting state.

\end{abstract}

\maketitle

Despite their discovery more than 30 years ago, there is still no consensus on the nature of the pseudogap (PG) phase of cuprate superconductors. This mysterious phase that emerges upon doping the parent Mott insulator shows many unusual features, the main one being a loss of density of states \cite{Alloul89,Warren89} as the temperature is decreased below a temperature $T^*$. In this study, we discuss a recent proposal where the opening of the pseudogap is attributed to the fractionalization of a Pair Density Wave (PDW) \cite{Chakraborty19}.  Within this framework, we discuss the complex phenomenology of Angle-Resolved Photo-Emission spectroscopy (ARPES) in nearly optimally doped Pb$_{0.55}$Bi$_{1.5}$Sr$_{1.6}$La$_{0.4}$CuO$_{6+\delta}$ (Bi2201). ARPES has proven to be one of the key probes in studying cuprates. It has notably been able to connect the loss of density of states in the pseudogap to the Fermi surface being gapped out in the antinodal region (ANR) for momenta close to $(0,\pm \pi),(\pm \pi,0)$, while other parts of the Fermi surface remain unchanged and form `Fermi arcs' (For a review of the pseudogap phenomenology see eg. \cite{Lee06,Norman03}).\\
\indent Most remarkably one observes, in the ANR, a back-bending at a momentum $k_G$ larger than the normal state Fermi momentum $k_F$, which suggested the presence of a modulation vector intimately linked with the opening of the pseudogap \cite{He11}. Moreover, as noticed in Ref \onlinecite{Lee14}, the gap continuously closes from `below' when moving towards the Fermi arcs, which is interpreted as revealing the presence of particle-particle pairs in the PG. Lastly, one observes that the bottom of the band at $k = (0,\pi)$ drops when decreasing the temperature and a new lightly dispersive `flat' band is seen in superconducting phase. Few other scenarios have been proposed to explain this specific gaping mechanism such as a quantum disorder PDW \cite{Dai18}, a coexistence of Charge Density Wave (CDW) and PDW \cite{Wang15b} or a Resonant Excitonic State (RES) \cite{Montiel:2016it}. In comparison to previous works, our study is based on a simple intuition, and accounts for all the experimental features with very few adjusting parameters.\\
\indent This paper is organized as follow. We start with a self-consistent equation for the pseudogap amplitude by treating a model of itinerant electrons interacting through antiferromagnetic exchange and residual density-density interaction at the mean-field level. The pseudogap can be seen as a superposition of superconductivity (SC) and CDW orders resulting in a composite order which has a non-zero amplitude in the ANR leading to the formation of Fermi arcs. The electronic spectral function in the ANR shows all the specific features mentioned previously, namely we obtain a back-bending of the electronic dispersion, a `flat' band and the gap closing from below when going closer to the center of the Brillouin zone. The effects of temperature on the spectral function close to $k = (\pi,0)$ and on the spectral weight of the flat band are also discussed for the first time based on phenomenological arguments.\\

The fractionalization of a PDW order is written in a way resembling the fractionalization of the electron introduced in strong coupling theories\cite{Baskaran88,Nagaosa90,Lee92}. The assumption is that, at a certain energy scale $E^{*}$, the system wants to form a PDW, which is depicted locally as an $\eta$-mode
\cite{Chakraborty19,Grandadam19}
\begin{align}
\hat{\eta} & =[\hat{\Delta}_{ij},\hat{\chi}_{ij}^{\dagger}], &  & \hat{\eta}^{\dagger}=[\hat{\chi}_{ij},\hat{\Delta}_{ij}^{\dagger}],\label{eq:1}
\end{align}
where $\hat{\Delta}_{ij}=\hat{d}_{ij}\sum_{\sigma}\sigma c_{i,\sigma}c_{j,-\sigma}$
and $\hat{\chi}_{ij}=\hat{d}_{ij}\sum_{\sigma}c_{i,\sigma}^{\dagger}c_{j,\sigma}e^{i\mathbf{Q.\left(\mathbf{r}_{i}+\mathbf{r}_{j}\right)/2}}$ are respectively the SC and CDW operators, $\hat{d}_{ij}$ being a structure factor which can assume d-wave symmetry and $\bm{Q}$ is the modulation wave-vector of the PDW. The $\eta$-operators are invariant with the following gauge structure
\begin{align}
\hat{\Delta}_{ij} & \rightarrow e^{i\theta}\hat{\Delta}_{ij}, &  & \hat{\chi}_{ij}\rightarrow e^{i\theta}\hat{\chi}_{ij}.\label{eq:2}
\end{align}
Then, the effective field theory for the fluctuating PDW is a rotor model\cite{sup}, in which the fluctuation of the U(1) gauge field produces a constraint between the two fields:
\begin{align}
 & |\hat{\Delta}_{ij}|^{2}+|\hat{\chi}_{ij}|^{2}\equiv\left|\Psi_{ij}\right|^{2}=const,\label{eq:frac}
\end{align}
where $\Psi_{ij}=(\hat{\Delta}_{ij},\hat{\chi}_{ij})^t$.
The energy scale associated to Eq.(\ref{eq:frac}) is typically the scale at which the fractionalization occurs. In our Ansatz, it corresponds to the PG scale which we denote $\left|\Psi_{ij}\right|=E^{*}$\cite{Chakraborty19,sup} in analogy with the pseudogap temperature $T^*$.

The goal of this paper is to test the validity of this unusual proposal by studying the fine structure of the spectral weight against the observations made by ARPES in Bi2201. For this we start with electrons hopping on a square lattice interacting
via an effective antiferromagnetic coupling $J_{ij}$, which comes for example from the Anderson super-exchange mechanism, and a small off-site residual Coulomb interaction term $V_{ij}$:
\begin{equation}
H=\sum_{i,j,\sigma}t_{ij}\left(c_{i\sigma}^{\dagger}c_{j\sigma}+h.c\right)+J_{ij}\ \bm{S}_{i}\cdot\bm{S}_{j}+V_{ij}\ n_{i}n_{j},\label{eq:H}
\end{equation}
where $c_{i\sigma}^{\dagger}$ is the creation operator for an electron
with spin $\sigma$ on a site $i$, $\bm{S_i} = c^{\dagger}_{i \alpha}\ \bm{\sigma}_{\alpha \beta}\ c_{i\beta}$ is the spin operator with $\bm{\sigma}$ the vector of Pauli matrices and $t_{ij}$ describe hopping between
different sites and are taken from a fit to ARPES data \cite{He11,Montiel:2016it,sup}. Both interactions are restricted to nearest-neighbours and we take $V$ to be smaller than $J$ which will be our main energy scale. \\

\begin{figure}[t]
\includegraphics[width = 8.6 cm]{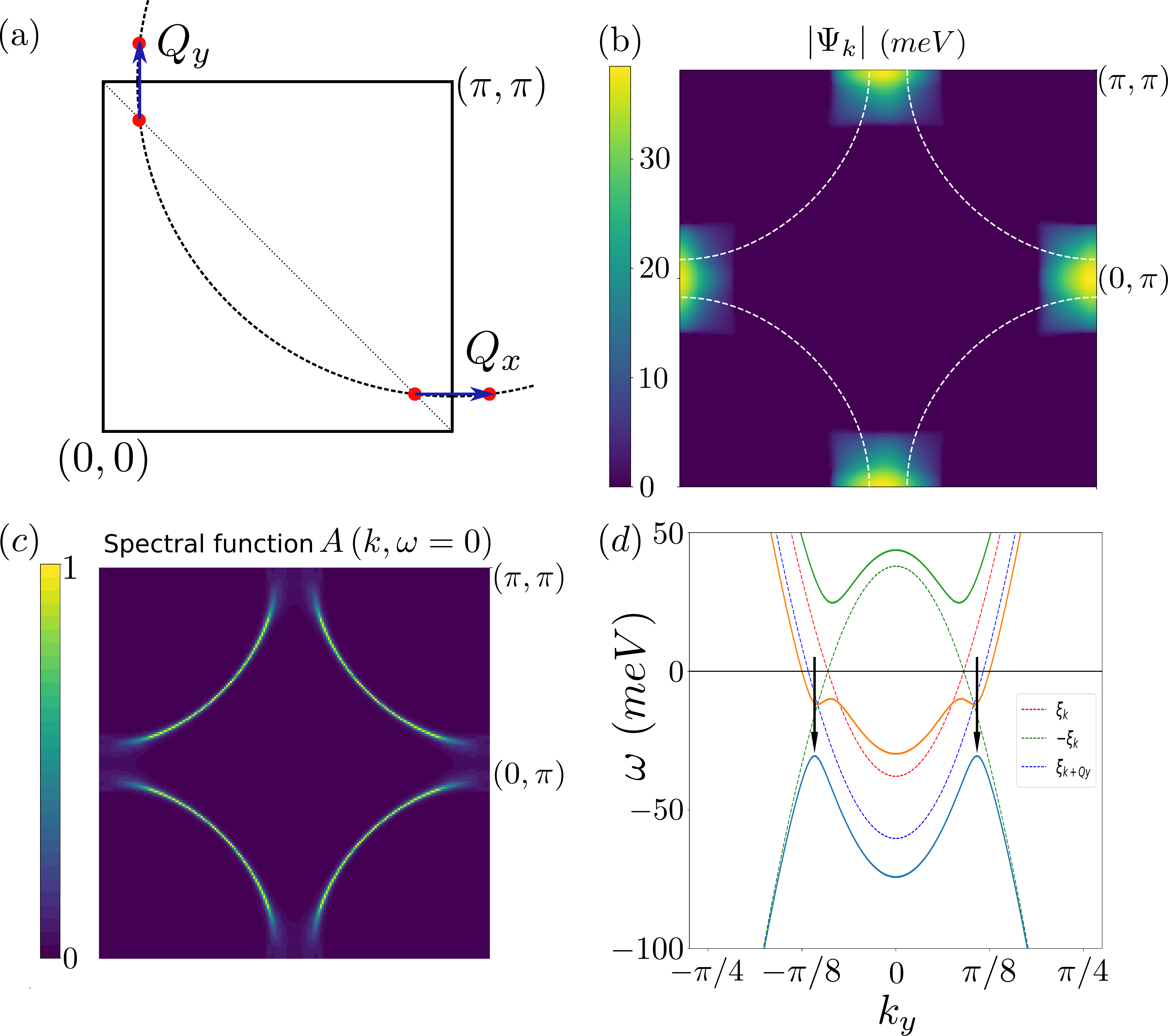}
\caption{\textbf{(a)} Schematic representation of the Brillouin zone for cuprates with the modulation wave-vector used in this work. \textbf{(b)} Solution of the gap equation for the pseudogap amplitude Eq.\eqref{eq:gap_eq}. Colored regions show non-zero solutions for $|\Psi_k|$. We used an axial modulation wave-vector $Q_x$ relating hot-spots shown in panel (a) and $J = 300\ meV$, $V = J/10$ and $q_{AF} = 0.15\ r.l.u$. The white line indicate the non-interacting Fermi surface. \textbf{(c)} Electronic spectral function $A\left( k,\omega = 0 \right)$ obtained from Eq.\eqref{eq:G} for $|\Psi_k|$ given by Eq.\eqref{eq:gap_eq}. We see the formation of Fermi arcs as the ANR gets gaped out. We used a broadening factor $\eta = 5\ meV$ for numerical purposes.  \textbf{(d)} Band structure obtain by the Green's function in Eq.\eqref{eq:G} for $k_x = \pi$ and a constant $|\Psi_k| = 30\ meV$. The dotted lines indicate the non-interacting dispersion $\xi_k$ (red), the hole band $-\xi_k$ (green) and the band from the modulating order $\xi_{k+Q_x}$ (blue). The black arrows point to the back-bending mentioned in the main text.}
\label{fig:fig1}
\end{figure}

\indent We treat this model in momentum space and start by decoupling the interaction with the individual fields forming the doublet. Here these fields are a pairing field $\Delta_{k}$ and four density modulation fields $\chi_{k}$ with uniaxial modulation vectors $Q = \pm Q_x; \pm Q_y = \left(\pm Q_0,0 \right);\left(0, \pm Q_0 \right)$ shown in Fig.\ref{fig:fig1}(a). The effective action for the doublet ${\Psi_k}^{\dagger} = \left( \Delta^*_k,\ {\chi_k}^* \right)$ representing our pseudogap is then given by\cite{sup} 
\begin{align}
&\mathcal{S}_{eff} = \int d \tau \sum_{Q,k,q} \frac{{\Psi_k}^{\dagger} \Psi_{k+q}}{\tilde{J}\left(q\right)} - \text{Tr}\ ln{\left(G^{-1}\left(i \omega, k\right)\right)},\label{eq:S} \\
&G^{-1} \left( i \omega,k \right) = i \omega -\xi_k - \sum_{Q = \pm Q_x, \pm Q_y}\frac{|\Psi_k|^2}{2} \tilde{G} \left( i\omega,k \right), \label{eq:G}
\end{align}
where $\xi_k$ is the non-interacting electronic dispersion, $\tilde{G}\left(i \omega,k \right) = \left(i \omega - \xi_{k+Q} \right)^{-1} + \left(i \omega + \xi_k \right)^{-1}$ and $1/\tilde{J} = 3J/\left(9J^2-V^2\right) $.
The electronic Green function in Eq.\eqref{eq:G} will thus be modified if $|\Psi_k|$ acquires a non zero value even if both the composing fields are fluctuating and have a vanishing expectation value.

\begin{figure}[t]
\includegraphics[width = 8.6 cm]{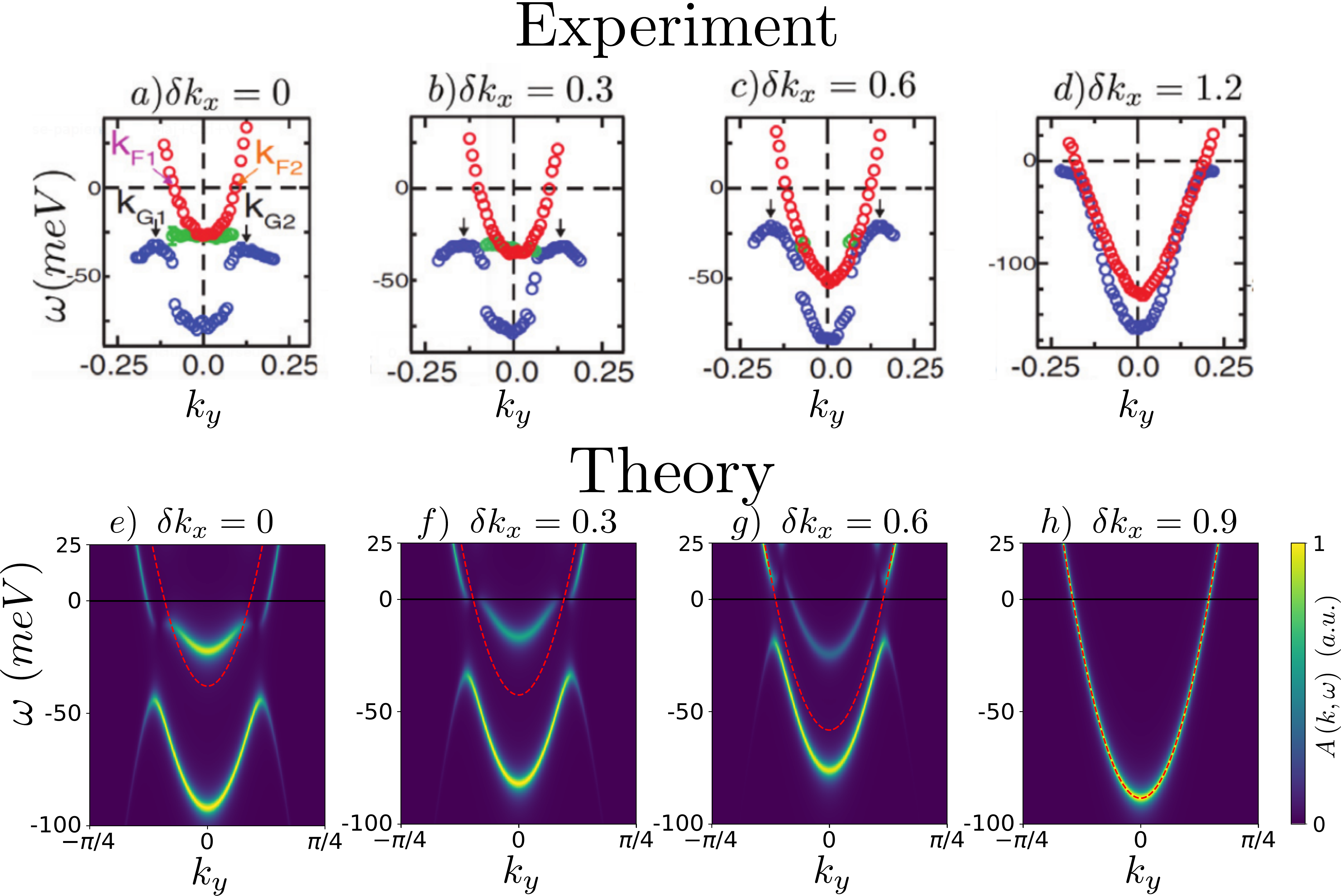}
\caption{\textbf{(a)-(d)} Experimental dispersion obtained by ARPES \cite{He11} for $T > T^*$ (red dots) and $T < T_c$ (blue and green dots) for different cuts at fixed $k_x = \pi - \delta k_x$. The Fermi arcs end around $\delta k_x = 0.6$ and the gap observed in the last panel is the standard nodal d-wave SC. \textbf{(e)-(h)} Theoretical results for the energy dependence of the spectral function $A \left( k,\omega \right)$ for cuts at fixed $k_x = \pi - \delta k_x$. The red dotted line is the non-interacting dispersion.  We used the solution of Eq.\eqref{eq:gap_eq} for the pseudogap amplitude and $Q_x = (0.2,0)\pi$}
\label{fig:fig2}
\end{figure}

\begin{figure}[t]
\includegraphics[width = 8.6 cm]{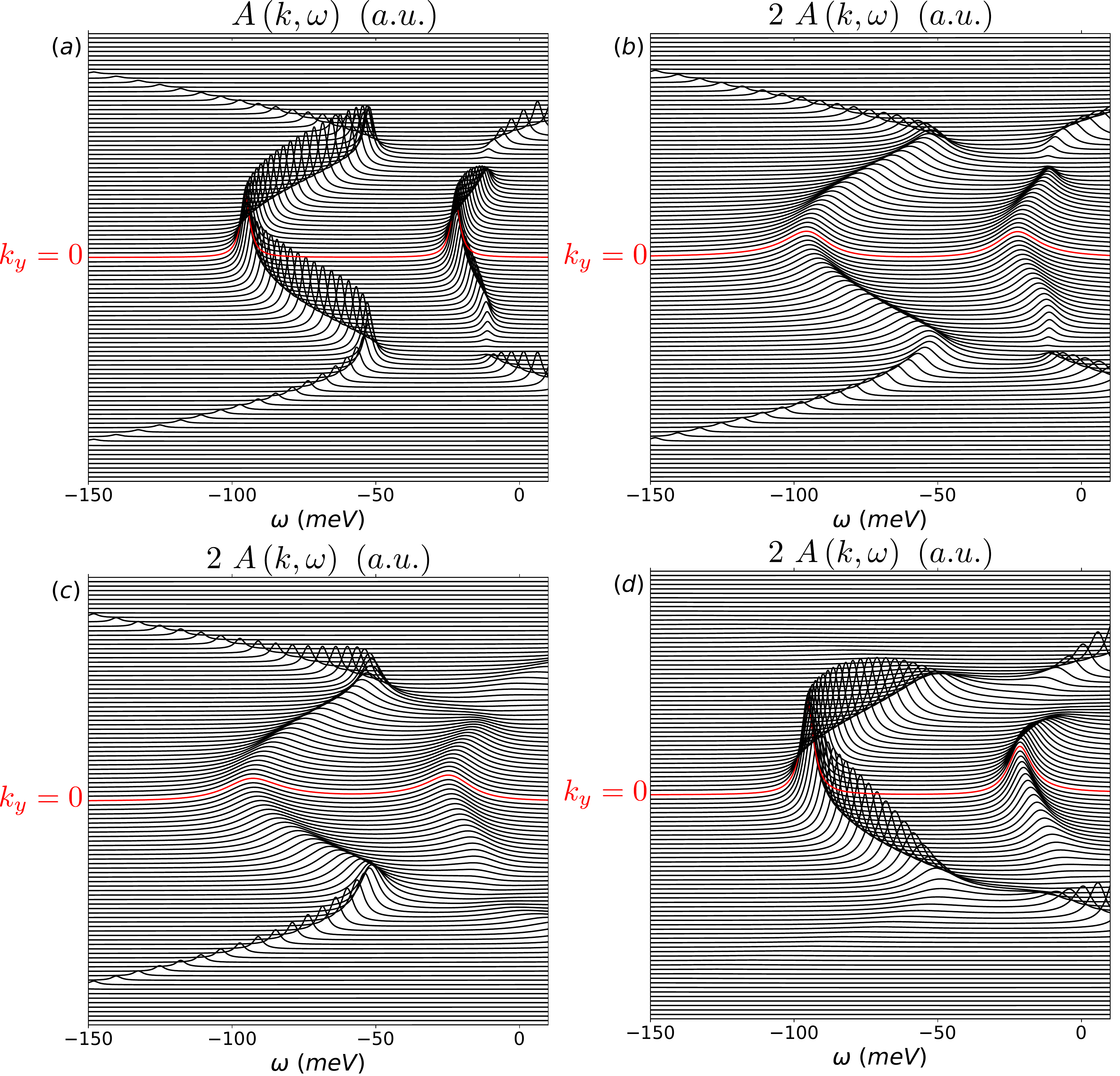}
\caption{Energy dependence of the spectral function at $k_x = \pi$ for different $k_y$ between $-\pi/4$ and $\pi/4$, successive lines are shifted for clarity. \textbf{(a)} without any lifetime $\Gamma_0 = \Gamma_1 = \Gamma_2 = 0 $. We used a broadening $\eta = 0.002\ eV$ for numerical purposes. \textbf{(b)} Turning on a single-particle lifetime $\Gamma_0 = 0.02\ eV$ leads to both bands being broadened in a similar way. \textbf{(c)} In contrast, when we consider only a finite particle-hole lifetime $\Gamma_1 = 0.02\ eV$ we see that the dispersion close to $k_y = 0$ is more strongly affected than the dispersion at higher momenta. The flat band is also more strongly dampened than the main band. \textbf{(d)} The situation is reversed if we consider only a particle-particle lifetime $\Gamma_2 = 0.02\ eV$. The parts of the bands close to $k_y = 0$ are less affected and still well defined. The flat band is more broadened but remains visible.}
\label{fig:fig3}
\end{figure}

\begin{figure}[t]
\includegraphics[width = 8.6 cm]{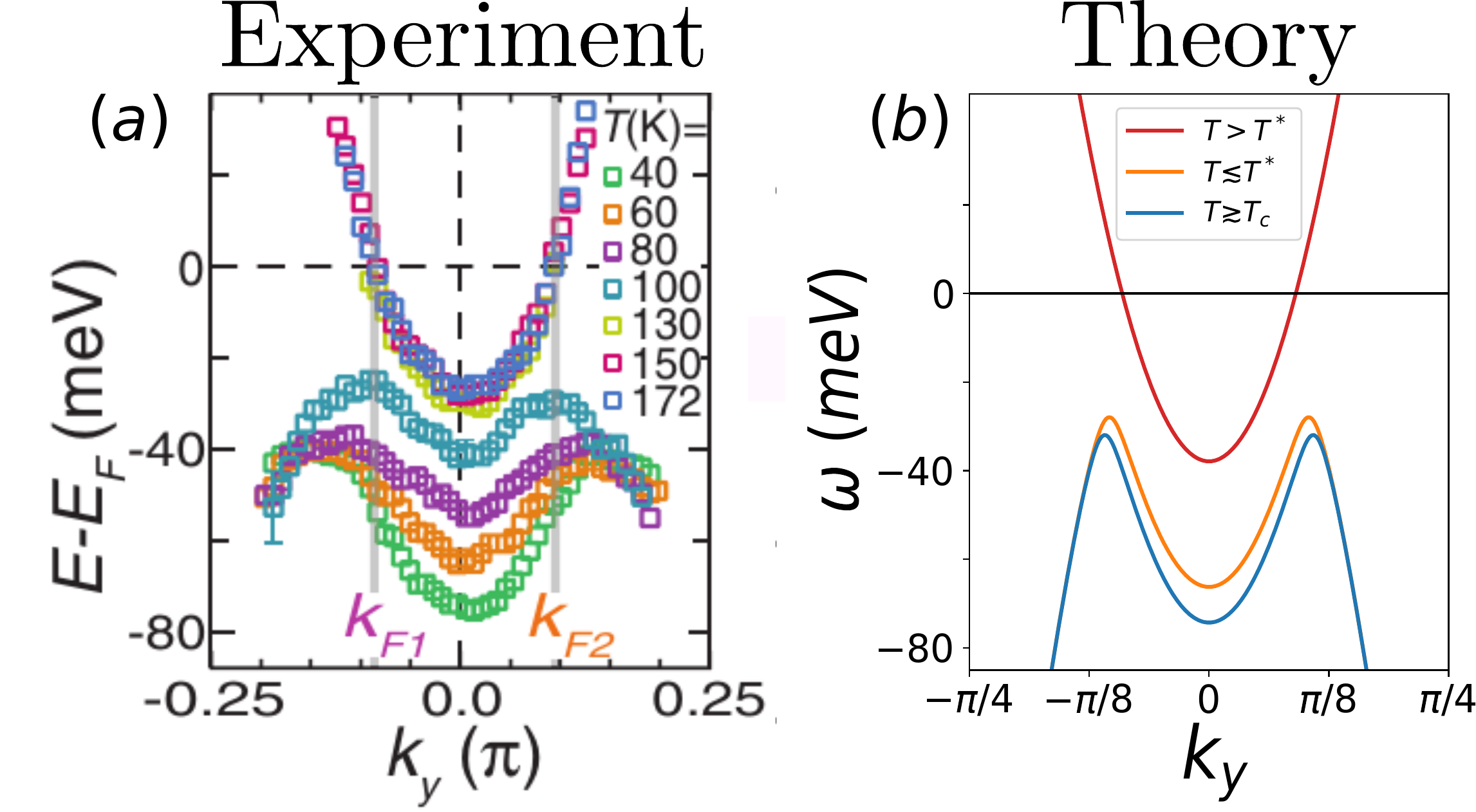}
\caption{Temperature evolution of the band at the zone edge in the pseudogap regime. \textbf{(a)} Experimental measurement for a range of temperature going from above $T^* \sim 132\ K$ to $T \gtrsim T_c \sim 38\ K$ \cite{He11} . \textbf{(b)}  The red line indicates the non-interacting band above $T^*$. The orange line is the band after the opening of the pseudogap presenting a back-bending shifted from the original Fermi momentum $k_F$. When going down in temperature we add a finite mean-field amplitude for the CDW order and obtain the band dispersion represented in blue. The back-bending wave-vector and the gap with respect to the Fermi level are mainly unchanged while the bottom of the band is strongly affected.}
\label{fig:fig4}
\end{figure}

Minimizing $\mathcal{S}_{eff}$ in Eq.\eqref{eq:S} with respect to the doublet $\Psi_k$ gives the mean-field gap equation for the doublet amplitude. We will consider the different modulation wave-vectors to be decoupled and use the fact that $\tilde{J}_{\bm{q}}$ is peaked around $\bm{q} = \left( \pi,\pi \right)$ to restrict the momentum summation to a range $q_{AF}$ around the antiferromagnetic wave-vector. The parameter $q_{AF}$ is physically associated with the short-range nature of the antiferromagnetic fluctuations \cite{Hinkov07} mediating the interaction. Assuming that $|\Psi_k|$ is constant over a range $q_{AF}$, the self-consistent equation is of the BCS form :
{\footnotesize \begin{align}
|\Psi_k| &= -\frac{T}{N}\sum_{i \omega_n, \tilde{q}} \tilde{J} \left(i\omega_n+\frac{\Delta \xi_{k+\tilde{q}}}{2}\right) \nonumber \\
& \times \frac{|\Psi_{k+\pi}|}{\left(\left(i\omega_n\right)^2-\xi_{k+\tilde{q}}^2\right) \left(i\omega_n-\xi_{k+Q+\tilde{q}}\right)-\left(i\omega_n+\frac{\Delta \xi_{k+\tilde{q}}}{2}\right)|\Psi_{k+\pi}|^2}\label{eq:gap_eq}
\end{align}}
where $\tilde{q}$ range between $\pi-q_{AF}/2$ and $\pi+q_{AF}/2$ and $\Delta \xi_{k+q} = \xi_{k+q}-\xi_{k+Q+q}$. Ignoring frequency dependence of $|\Psi_k|$ we can perform the Matsubara summation analytically. This leads to two coupled equation between $|\Psi_k|$ and $|\Psi_{k+\bm{\pi}}|$ which we can solve self-consistently. \\\\
\indent Results of this self-consistent equation are shown in Fig.\ref{fig:fig1}(b) for a modulation wave-vector linking hot-spots along the $x$ axis. Hot-spots are points of the Fermi surface linked by $\left(\pi,\pi \right)$ and are thus expected to be important due to the form of our interaction. Due to the finite wave-vector of the pseudogap amplitude, the gap equation Eq.\eqref{eq:gap_eq} admits non-zero solution only in the ANR, when this modulation vector links two parts of the Fermi surface. The region close to the Brillouin zone diagonal will thus remain unperturbed by the transition at $T^*$. The electronic spectral function $A\left(\omega,k\right) = -\frac{1}{\pi} Im\left( G\left(\omega + i\eta,k\right)\right)$ for $\omega = 0$ and $\eta \rightarrow 0^+$ shows that the ANR is gapped while the nodal region forms Fermi arcs (Fig.\ref{fig:fig1}(c)). These arcs terminate close to the hot-spots.
\indent We now look at the reconstructed band structure obtained from the zeros of $G^{-1}(k,\omega)$ in the ANR. From the form of $\tilde{G}\left(k,\omega\right)$, we can understand the dispersion as coming from an equal superposition of SC and CDW order in the ANR. We can thus construct the resulting band structure in the pseudogap as the hybridization of the three bands $\xi_k$, the normal state dispersion, $-\xi_k$ coming from the superconducting order and $\xi_{k+Q}$ coming from the modulating order. At the zone boundary ($k_x = \pi$), this results in two bands below the Fermi level shown in Fig.\ref{fig:fig1}(d) with one of them presenting a back-bending (blue line) indicated by black arrows while the other one (yellow line) present little dispersion around $k_y = 0$. In our mean-field description, this back-bending appears as a result of the hybridization between the hole band $-\xi_k$ (green doted) and the shifted $\xi_{k+Q}$ (blue doted) band in Fig.\ref{fig:fig1}(d). As such this back-bending will occur at $k_y = k_G > k_F$ as long as $\xi_{k+Q} < \xi_k$ which is satisfied for all $k_x > k_{hot-spot}$. This means that this anomalous back-bending will persist below $T_c$ in the ANR but we will recover a standard back-bending at $k_y = k_F$ in the nodal region as the above condition is not satisfied. \\
\indent The spectral weight $A(k,\omega)$ for each band is obtained for different fixed values of $k_x = \pi-\delta k_x$ and compared to the experimental dispersion of Ref.\onlinecite{He11} (Fig.\ref{fig:fig2}(a)-(d)). As we get closer to the centre of the Brillouin zone we can see that the energy of the maximum of the band gets closer to the Fermi level leading to the pseudogap closing `from below' (Fig.\ref{fig:fig2}(e)-(h)) as observed experimentally. Note that we obtain here a gap closing `from below' contrary to what was argued previously for a pure CDW scenario with a modulation along the $y$ direction \cite{Lee14}. This is because we consider a modulation wave-vector along the $x$ direction. This same orientation for the modulation wave-vector has been used recently to explain ARPES results in Bi2201 through the idea of a quantum disorder PDW \cite{Dai18} and other theoretical approaches such as a superposition of CDW and PDW order \cite{Wang15b} or a RES \cite{Montiel:2016it}.
\\ \indent Our previous description of the band structure in the pseudogap also shows a second band located at the bottom of the non-interacting band. We connect here this band to the flat band observed experimentally below $T_c$ (green dots in Fig.\ref{fig:fig2}(a)-(c)) and argue that finite lifetimes for the single-particle and pair excitations lead to this band not being observed above $T_c$. For this, we add three phenomenological damping rates $\Gamma_0$, $\Gamma_1$ and $\Gamma_2$ in our mean-field Green's function :
\begin{align}
G^{-1} \left( i \omega,k \right) &= i \omega -\xi_k - i \Gamma_0 - \sum_{Q = \pm Q_x, \pm Q_y}\frac{|\Psi_k|^2}{2} \tilde{G} \left( i\omega,k \right), \nonumber \\
\tilde{G}\left(i \omega,k \right) &= \left(i \omega - \xi_{k+Q} + i \Gamma_1 \right)^{-1} + \left(i \omega + \xi_k + i \Gamma_2\right)^{-1}.
\end{align}
The two factors $\Gamma_1$ and $\Gamma_2$ represent the lifetime of particle-hole and particle-particle pairs respectively \cite{Norman:1995dd,Banerjee:2011cu,Banerjee:2011bz}. These lifetimes capture the fluctuations in the pseudogap phase and are used in other approaches such as preformed pairs\cite{Norman:1998va,Chien2009,Campuzano98,Campuzano:1996fb} or effect of gaussian fluctuations\cite{Benfatto00}. They are expected to be non-zero above $T_c$ but to vanish at the transition temperature when fluctuations are quenched. The first $\Gamma_0$ term is a single-particle lifetime which is always non-zero. The effect of each of these additional terms is depicted in Fig.\ref{fig:fig3}(a)-(d). Allowing a non-zero $\Gamma_0$ will broaden the two bands below the Fermi level in similar ways (Fig.\ref{fig:fig3}(b)), in contrast to the pair lifetimes which have a very different effect on specific parts of the dispersion. Indeed, Fig.\ref{fig:fig3}(c) shows that a non-zero $\Gamma_1$ will strongly suppress the `flat' band close to the Fermi level and also dampen the main band close to $k_y=0$. Turning on the $\Gamma_2$ term will have the opposite effect as the band far from $k_y=0$ gets dampened while the bottom of the flat and main bands remain well defined. The experimental observation of the flat band only close or below $T_c$ can then be attributed to the presence of a particle-hole pair lifetime in the pseudogap. Note also that due to disorder effect, which couple directly to charge order \cite{DelMaestro06}, this lifetime could remain non-zero below $T_c$ and thus leads to this band remaining broad even in the superconducting state as observed experimentally. Moreover, this description provides a good agreement with the experimental observation that the dispersion in the ANR does not change across the superconducting transition. In our case, the position of the main band does not change with temperature and only the spectral weights of the two bands get modified as the different lifetimes decrease.
\\
\indent Another feature of the temperature dependence measured experimentally for $T^* > T > T_c$ in the ANR is a significant decrease of the energy of bottom of the band when the temperature is decreased while the maximum energy and the back-bending wave-vector change only slightly as shown in Fig.\ref{fig:fig4}(a). We describe here this change in the band structure by adding a finite amplitude for the particle-hole order parameter $|\chi_k|$. We then have three different regions such that at $T>T^*$ we have free electrons. At $T \lesssim T^*$ where the pseudogap has a finite amplitude but the particle-hole gap is still 0 and at $T \gtrsim T_c$ where the particle-hole gap is finite. We then obtain the band dispersions shown in Fig.\ref{fig:fig4}(b). Because the bottom of the band is directly related to the hybridization with the band coming from the charge modulation, it is directly affected by the non-zero value of $|\chi_k|$. On the other hand, the back-bending momentum is determined mainly by the value of the modulation wave-vector $Q$ and the energy of the maximum comes from the hybridization between the superconducting band $-\xi_k$ and the shifted band $\xi_{k+Q}$, thus related to the value of $|\Psi_k|$. In contrast, a finite amplitude for the SC order parameter $|\Delta_k|$ would produce the opposite effect and change substantially the position of the maximum leaving the bottom of the band unchanged\cite{sup}. The fact that $|\chi_k|$ acquire a quasi long-range component before the $|\Delta_k|$ is representative of the fact that CDW is observed experimentally at a temperature higher than the temperature for SC fluctuations $T_c^{\prime}$. This long-range component of the CDW order has also been observed by Raman spectroscopy \cite{Loret19}, X-ray \cite{Chang12,Blanco-Canosa13,Blackburn13a,Ghiringhelli12,Gerber:2015gx,Chang16} and NMR \cite{Wu11,Wu13a,Wu:2015bt,Julien15} measurements above $T_c$.\\

In conclusion, we showed here how the recently proposed idea of fractionalized PDW \cite{Chakraborty19} can be used to construct a mean-field description of the pseudogap phase of cuprates. The main idea is that even if none of CDW or SC orders develop a long range-component, the constraint introduced by the fractionalization affects the electronic Green's function. Using a microscopic model we derived a self-consistent equation for the pseudogap amplitude and showed that it has non-zero solutions in the antinodal region, which gives a gap in the ANR in the PG phase and leads to the formation of Fermi arcs. The band dispersion obtained in the antinodal region is in good agreement with the experimental ARPES measurement made on Bi2201. Especially, we recover all the features observed in the pseudogap state. The superposition of particle-particle and particle-hole orders leads to an anomalous back-bending of the main band below the Fermi level, a gap closing `from below' and a `flat band' close to the bottom of the original electronic dispersion. We argue that this band is seen experimentally only below $T_c$ because it is strongly affected by the finite pair lifetime in the pseudogap phase. Lastly, we discussed the change of the dispersion as the temperature is lowered from $T^*$ to $T_c$ by showing that a finite quasi long-range component of the particle-hole order leads to the minimum of the band going down in energy while the energy and momentum of the maximum stay unchanged.\\
\indent The competition between different orders is present in many other materials such as transition metal dichalcogenides\cite{Koley20} for example. We showed here that considering an entanglement between these competing orders has unique consequences beyond the standard competing scenarios. This idea could also be used to study other materials that exhibit pseudogap physics such as CeRhIn$_5$\cite{Kawasaki05} and NbSe$_2$\cite{Borisenko09,Chatterjee15}.\\\\
We thank S. Sarkar and A. Banerjee for valuable discussions. This work has received financial support from the ERC, under grant agreement AdG-694651-CHAMPAGNE.

\bibliographystyle{apsrev4-1}
\bibliography{Cuprates}

\begin{thebibliography}{40}%
\makeatletter
\providecommand \@ifxundefined [1]{%
 \@ifx{#1\undefined}
}%
\providecommand \@ifnum [1]{%
 \ifnum #1\expandafter \@firstoftwo
 \else \expandafter \@secondoftwo
 \fi
}%
\providecommand \@ifx [1]{%
 \ifx #1\expandafter \@firstoftwo
 \else \expandafter \@secondoftwo
 \fi
}%
\providecommand \natexlab [1]{#1}%
\providecommand \enquote  [1]{``#1''}%
\providecommand \bibnamefont  [1]{#1}%
\providecommand \bibfnamefont [1]{#1}%
\providecommand \citenamefont [1]{#1}%
\providecommand \href@noop [0]{\@secondoftwo}%
\providecommand \href [0]{\begingroup \@sanitize@url \@href}%
\providecommand \@href[1]{\@@startlink{#1}\@@href}%
\providecommand \@@href[1]{\endgroup#1\@@endlink}%
\providecommand \@sanitize@url [0]{\catcode `\\12\catcode `\$12\catcode
  `\&12\catcode `\#12\catcode `\^12\catcode `\_12\catcode `\%12\relax}%
\providecommand \@@startlink[1]{}%
\providecommand \@@endlink[0]{}%
\providecommand \url  [0]{\begingroup\@sanitize@url \@url }%
\providecommand \@url [1]{\endgroup\@href {#1}{\urlprefix }}%
\providecommand \urlprefix  [0]{URL }%
\providecommand \Eprint [0]{\href }%
\providecommand \doibase [0]{http://dx.doi.org/}%
\providecommand \selectlanguage [0]{\@gobble}%
\providecommand \bibinfo  [0]{\@secondoftwo}%
\providecommand \bibfield  [0]{\@secondoftwo}%
\providecommand \translation [1]{[#1]}%
\providecommand \BibitemOpen [0]{}%
\providecommand \bibitemStop [0]{}%
\providecommand \bibitemNoStop [0]{.\EOS\space}%
\providecommand \EOS [0]{\spacefactor3000\relax}%
\providecommand \BibitemShut  [1]{\csname bibitem#1\endcsname}%
\let\auto@bib@innerbib\@empty
\bibitem [{\citenamefont {Alloul}\ \emph {et~al.}(1989)\citenamefont {Alloul},
  \citenamefont {Ohno},\ and\ \citenamefont {Mendels}}]{Alloul89}%
  \BibitemOpen
  \bibfield  {author} {\bibinfo {author} {\bibfnamefont {H.}~\bibnamefont
  {Alloul}}, \bibinfo {author} {\bibfnamefont {T.}~\bibnamefont {Ohno}}, \ and\
  \bibinfo {author} {\bibfnamefont {P.}~\bibnamefont {Mendels}},\ }\href
  {\doibase 10.1103/PhysRevLett.63.1700} {\bibfield  {journal} {\bibinfo
  {journal} {Phys. Rev. Lett.}\ }\textbf {\bibinfo {volume} {63}},\ \bibinfo
  {pages} {1700} (\bibinfo {year} {1989})}\BibitemShut {NoStop}%
\bibitem [{\citenamefont {Warren}\ \emph {et~al.}(1989)\citenamefont {Warren},
  \citenamefont {Walstedt}, \citenamefont {Brennert}, \citenamefont {Cava},
  \citenamefont {Tycko}, \citenamefont {Bell},\ and\ \citenamefont
  {Dabbagh}}]{Warren89}%
  \BibitemOpen
  \bibfield  {author} {\bibinfo {author} {\bibfnamefont {W.~W.}\ \bibnamefont
  {Warren}}, \bibinfo {author} {\bibfnamefont {R.~E.}\ \bibnamefont
  {Walstedt}}, \bibinfo {author} {\bibfnamefont {G.~F.}\ \bibnamefont
  {Brennert}}, \bibinfo {author} {\bibfnamefont {R.~J.}\ \bibnamefont {Cava}},
  \bibinfo {author} {\bibfnamefont {R.}~\bibnamefont {Tycko}}, \bibinfo
  {author} {\bibfnamefont {R.~F.}\ \bibnamefont {Bell}}, \ and\ \bibinfo
  {author} {\bibfnamefont {G.}~\bibnamefont {Dabbagh}},\ }\href {\doibase
  10.1103/PhysRevLett.62.1193} {\bibfield  {journal} {\bibinfo  {journal}
  {Phys. Rev. Lett.}\ }\textbf {\bibinfo {volume} {62}},\ \bibinfo {pages}
  {1193} (\bibinfo {year} {1989})}\BibitemShut {NoStop}%
\bibitem [{\citenamefont {{Chakraborty}}\ \emph {et~al.}(2019)\citenamefont
  {{Chakraborty}}, \citenamefont {{Grandadam}}, \citenamefont {{Hamidian}},
  \citenamefont {{Davis}}, \citenamefont {{Sidis}},\ and\ \citenamefont
  {{P{\'e}pin}}}]{Chakraborty19}%
  \BibitemOpen
  \bibfield  {author} {\bibinfo {author} {\bibfnamefont {D.}~\bibnamefont
  {{Chakraborty}}}, \bibinfo {author} {\bibfnamefont {M.}~\bibnamefont
  {{Grandadam}}}, \bibinfo {author} {\bibfnamefont {M.~H.}\ \bibnamefont
  {{Hamidian}}}, \bibinfo {author} {\bibfnamefont {J.~C.~S.}\ \bibnamefont
  {{Davis}}}, \bibinfo {author} {\bibfnamefont {Y.}~\bibnamefont {{Sidis}}}, \
  and\ \bibinfo {author} {\bibfnamefont {C.}~\bibnamefont {{P{\'e}pin}}},\
  }\href@noop {} {\bibfield  {journal} {\bibinfo  {journal} {arXiv e-prints}\
  ,\ \bibinfo {eid} {arXiv:1906.01633}} (\bibinfo {year} {2019})},\ \Eprint
  {http://arxiv.org/abs/1906.01633} {arXiv:1906.01633 [cond-mat.supr-con]}
  \BibitemShut {NoStop}%
\bibitem [{\citenamefont {Lee}\ \emph {et~al.}(2006)\citenamefont {Lee},
  \citenamefont {Nagaosa},\ and\ \citenamefont {Wen}}]{Lee06}%
  \BibitemOpen
  \bibfield  {author} {\bibinfo {author} {\bibfnamefont {P.~A.}\ \bibnamefont
  {Lee}}, \bibinfo {author} {\bibfnamefont {N.}~\bibnamefont {Nagaosa}}, \ and\
  \bibinfo {author} {\bibfnamefont {X.-G.}\ \bibnamefont {Wen}},\ }\href
  {\doibase 10.1103/RevModPhys.78.17} {\bibfield  {journal} {\bibinfo
  {journal} {Rev. Mod. Phys.}\ }\textbf {\bibinfo {volume} {78}},\ \bibinfo
  {pages} {17} (\bibinfo {year} {2006})}\BibitemShut {NoStop}%
\bibitem [{\citenamefont {Norman}\ and\ \citenamefont
  {P\'epin}(2003)}]{Norman03}%
  \BibitemOpen
  \bibfield  {author} {\bibinfo {author} {\bibfnamefont {M.~R.}\ \bibnamefont
  {Norman}}\ and\ \bibinfo {author} {\bibfnamefont {C.}~\bibnamefont
  {P\'epin}},\ }\href {http://stacks.iop.org/0034-4885/66/i=10/a=R01}
  {\bibfield  {journal} {\bibinfo  {journal} {Rep. Prog. Phys.}\ }\textbf
  {\bibinfo {volume} {66}},\ \bibinfo {pages} {1547} (\bibinfo {year}
  {2003})}\BibitemShut {NoStop}%
\bibitem [{\citenamefont {He}\ \emph {et~al.}(2011)\citenamefont {He},
  \citenamefont {Hashimoto}, \citenamefont {Karapetyan}, \citenamefont
  {Koralek}, \citenamefont {Hinton}, \citenamefont {Testaud}, \citenamefont
  {Nathan}, \citenamefont {Yoshida}, \citenamefont {Yao}, \citenamefont
  {Tanaka}, \citenamefont {Meevasana}, \citenamefont {Moore}, \citenamefont
  {Lu}, \citenamefont {Mo}, \citenamefont {Ishikado}, \citenamefont {Eisaki},
  \citenamefont {Hussain}, \citenamefont {Devereaux}, \citenamefont {Kivelson},
  \citenamefont {Orenstein}, \citenamefont {Kapitulnik},\ and\ \citenamefont
  {Shen}}]{He11}%
  \BibitemOpen
  \bibfield  {author} {\bibinfo {author} {\bibfnamefont {R.-H.}\ \bibnamefont
  {He}}, \bibinfo {author} {\bibfnamefont {M.}~\bibnamefont {Hashimoto}},
  \bibinfo {author} {\bibfnamefont {H.}~\bibnamefont {Karapetyan}}, \bibinfo
  {author} {\bibfnamefont {J.~D.}\ \bibnamefont {Koralek}}, \bibinfo {author}
  {\bibfnamefont {J.~P.}\ \bibnamefont {Hinton}}, \bibinfo {author}
  {\bibfnamefont {J.~P.}\ \bibnamefont {Testaud}}, \bibinfo {author}
  {\bibfnamefont {V.}~\bibnamefont {Nathan}}, \bibinfo {author} {\bibfnamefont
  {Y.}~\bibnamefont {Yoshida}}, \bibinfo {author} {\bibfnamefont
  {H.}~\bibnamefont {Yao}}, \bibinfo {author} {\bibfnamefont {K.}~\bibnamefont
  {Tanaka}}, \bibinfo {author} {\bibfnamefont {W.}~\bibnamefont {Meevasana}},
  \bibinfo {author} {\bibfnamefont {R.~G.}\ \bibnamefont {Moore}}, \bibinfo
  {author} {\bibfnamefont {D.~H.}\ \bibnamefont {Lu}}, \bibinfo {author}
  {\bibfnamefont {S.-K.}\ \bibnamefont {Mo}}, \bibinfo {author} {\bibfnamefont
  {M.}~\bibnamefont {Ishikado}}, \bibinfo {author} {\bibfnamefont
  {H.}~\bibnamefont {Eisaki}}, \bibinfo {author} {\bibfnamefont
  {Z.}~\bibnamefont {Hussain}}, \bibinfo {author} {\bibfnamefont {T.~P.}\
  \bibnamefont {Devereaux}}, \bibinfo {author} {\bibfnamefont {S.~A.}\
  \bibnamefont {Kivelson}}, \bibinfo {author} {\bibfnamefont {J.}~\bibnamefont
  {Orenstein}}, \bibinfo {author} {\bibfnamefont {A.}~\bibnamefont
  {Kapitulnik}}, \ and\ \bibinfo {author} {\bibfnamefont {Z.-X.}\ \bibnamefont
  {Shen}},\ }\href {\doibase 10.1126/science.1198415} {\bibfield  {journal}
  {\bibinfo  {journal} {Science}\ }\textbf {\bibinfo {volume} {331}},\ \bibinfo
  {pages} {1579} (\bibinfo {year} {2011})}\BibitemShut {NoStop}%
\bibitem [{\citenamefont {Lee}(2014)}]{Lee14}%
  \BibitemOpen
  \bibfield  {author} {\bibinfo {author} {\bibfnamefont {P.~A.}\ \bibnamefont
  {Lee}},\ }\href {\doibase 10.1103/PhysRevX.4.031017} {\bibfield  {journal}
  {\bibinfo  {journal} {Phys. Rev. X}\ }\textbf {\bibinfo {volume} {4}},\
  \bibinfo {pages} {031017} (\bibinfo {year} {2014})}\BibitemShut {NoStop}%
\bibitem [{\citenamefont {Dai}\ \emph {et~al.}(2018)\citenamefont {Dai},
  \citenamefont {Zhang}, \citenamefont {Senthil},\ and\ \citenamefont
  {Lee}}]{Dai18}%
  \BibitemOpen
  \bibfield  {author} {\bibinfo {author} {\bibfnamefont {Z.}~\bibnamefont
  {Dai}}, \bibinfo {author} {\bibfnamefont {Y.-H.}\ \bibnamefont {Zhang}},
  \bibinfo {author} {\bibfnamefont {T.}~\bibnamefont {Senthil}}, \ and\
  \bibinfo {author} {\bibfnamefont {P.~A.}\ \bibnamefont {Lee}},\ }\href
  {\doibase 10.1103/PhysRevB.97.174511} {\bibfield  {journal} {\bibinfo
  {journal} {Phys. Rev. B}\ }\textbf {\bibinfo {volume} {97}},\ \bibinfo
  {pages} {174511} (\bibinfo {year} {2018})}\BibitemShut {NoStop}%
\bibitem [{\citenamefont {Wang}\ \emph {et~al.}(2015)\citenamefont {Wang},
  \citenamefont {Agterberg},\ and\ \citenamefont {Chubukov}}]{Wang15b}%
  \BibitemOpen
  \bibfield  {author} {\bibinfo {author} {\bibfnamefont {Y.}~\bibnamefont
  {Wang}}, \bibinfo {author} {\bibfnamefont {D.~F.}\ \bibnamefont {Agterberg}},
  \ and\ \bibinfo {author} {\bibfnamefont {A.}~\bibnamefont {Chubukov}},\
  }\href {\doibase 10.1103/PhysRevLett.114.197001} {\bibfield  {journal}
  {\bibinfo  {journal} {Phys. Rev. Lett.}\ }\textbf {\bibinfo {volume} {114}},\
  \bibinfo {pages} {197001} (\bibinfo {year} {2015})}\BibitemShut {NoStop}%
\bibitem [{\citenamefont {Montiel}\ \emph {et~al.}(2016)\citenamefont
  {Montiel}, \citenamefont {Kloss},\ and\ \citenamefont
  {P{\'e}pin}}]{Montiel:2016it}%
  \BibitemOpen
  \bibfield  {author} {\bibinfo {author} {\bibfnamefont {X.}~\bibnamefont
  {Montiel}}, \bibinfo {author} {\bibfnamefont {T.}~\bibnamefont {Kloss}}, \
  and\ \bibinfo {author} {\bibfnamefont {C.}~\bibnamefont {P{\'e}pin}},\ }\href
  {\doibase 10.1209/0295-5075/115/57001} {\bibfield  {journal} {\bibinfo
  {journal} {EPL (Europhysics Letters)}\ }\textbf {\bibinfo {volume} {115}},\
  \bibinfo {pages} {57001} (\bibinfo {year} {2016})}\BibitemShut {NoStop}%
\bibitem [{\citenamefont {Baskaran}\ and\ \citenamefont
  {Anderson}(1988)}]{Baskaran88}%
  \BibitemOpen
  \bibfield  {author} {\bibinfo {author} {\bibfnamefont {G.}~\bibnamefont
  {Baskaran}}\ and\ \bibinfo {author} {\bibfnamefont {P.~W.}\ \bibnamefont
  {Anderson}},\ }\href {\doibase 10.1103/PhysRevB.37.580} {\bibfield  {journal}
  {\bibinfo  {journal} {Phys. Rev. B}\ }\textbf {\bibinfo {volume} {37}},\
  \bibinfo {pages} {580} (\bibinfo {year} {1988})}\BibitemShut {NoStop}%
\bibitem [{\citenamefont {Nagaosa}\ and\ \citenamefont
  {Lee}(1990)}]{Nagaosa90}%
  \BibitemOpen
  \bibfield  {author} {\bibinfo {author} {\bibfnamefont {N.}~\bibnamefont
  {Nagaosa}}\ and\ \bibinfo {author} {\bibfnamefont {P.~A.}\ \bibnamefont
  {Lee}},\ }\href {\doibase 10.1103/PhysRevLett.64.2450} {\bibfield  {journal}
  {\bibinfo  {journal} {Phys. Rev. Lett.}\ }\textbf {\bibinfo {volume} {64}},\
  \bibinfo {pages} {2450} (\bibinfo {year} {1990})}\BibitemShut {NoStop}%
\bibitem [{\citenamefont {Lee}\ and\ \citenamefont {Nagaosa}(1992)}]{Lee92}%
  \BibitemOpen
  \bibfield  {author} {\bibinfo {author} {\bibfnamefont {P.~A.}\ \bibnamefont
  {Lee}}\ and\ \bibinfo {author} {\bibfnamefont {N.}~\bibnamefont {Nagaosa}},\
  }\href {\doibase 10.1103/PhysRevB.46.5621} {\bibfield  {journal} {\bibinfo
  {journal} {Phys. Rev. B}\ }\textbf {\bibinfo {volume} {46}},\ \bibinfo
  {pages} {5621} (\bibinfo {year} {1992})}\BibitemShut {NoStop}%
\bibitem [{\citenamefont {{Grandadam}}\ \emph {et~al.}(2019)\citenamefont
  {{Grandadam}}, \citenamefont {{Chakraborty}},\ and\ \citenamefont
  {{P{\'e}pin}}}]{Grandadam19}%
  \BibitemOpen
  \bibfield  {author} {\bibinfo {author} {\bibfnamefont {M.}~\bibnamefont
  {{Grandadam}}}, \bibinfo {author} {\bibfnamefont {D.}~\bibnamefont
  {{Chakraborty}}}, \ and\ \bibinfo {author} {\bibfnamefont {C.}~\bibnamefont
  {{P{\'e}pin}}},\ }\href@noop {} {\bibfield  {journal} {\bibinfo  {journal}
  {arXiv e-prints}\ ,\ \bibinfo {eid} {arXiv:1909.06657}} (\bibinfo {year}
  {2019})},\ \Eprint {http://arxiv.org/abs/1909.06657} {arXiv:1909.06657
  [cond-mat.supr-con]} \BibitemShut {NoStop}%
\bibitem [{\citenamefont {Grandadam}\ \emph {et~al.}()\citenamefont
  {Grandadam}, \citenamefont {Chakraborty}, \citenamefont {Montiel},\ and\
  \citenamefont {P{\'e}pin}}]{sup}%
  \BibitemOpen
  \bibfield  {author} {\bibinfo {author} {\bibfnamefont {M.}~\bibnamefont
  {Grandadam}}, \bibinfo {author} {\bibfnamefont {D.}~\bibnamefont
  {Chakraborty}}, \bibinfo {author} {\bibfnamefont {X.}~\bibnamefont
  {Montiel}}, \ and\ \bibinfo {author} {\bibfnamefont {C.}~\bibnamefont
  {P{\'e}pin}},\ }\href@noop {} {\bibinfo  {journal} {Supplementary Materials}\
  }\BibitemShut {NoStop}%
\bibitem [{\citenamefont {Hinkov}\ \emph {et~al.}(2007)\citenamefont {Hinkov},
  \citenamefont {Bourges}, \citenamefont {Pailh\`{e}s}, \citenamefont {Ivanov},
  \citenamefont {Frost}, \citenamefont {Perring}, \citenamefont {Lin},
  \citenamefont {Chen},\ and\ \citenamefont {Keimer}}]{Hinkov07}%
  \BibitemOpen
\bibfield  {journal} {  }\bibfield  {author} {\bibinfo {author} {\bibfnamefont
  {V.}~\bibnamefont {Hinkov}}, \bibinfo {author} {\bibfnamefont
  {P.}~\bibnamefont {Bourges}}, \bibinfo {author} {\bibfnamefont
  {Y.}~\bibnamefont {Pailh\`{e}s}, \bibfnamefont {S.and~Sidis}}, \bibinfo
  {author} {\bibfnamefont {A.}~\bibnamefont {Ivanov}}, \bibinfo {author}
  {\bibfnamefont {C.~D.}\ \bibnamefont {Frost}}, \bibinfo {author}
  {\bibfnamefont {T.~G.}\ \bibnamefont {Perring}}, \bibinfo {author}
  {\bibfnamefont {C.~T.}\ \bibnamefont {Lin}}, \bibinfo {author} {\bibfnamefont
  {D.~P.}\ \bibnamefont {Chen}}, \ and\ \bibinfo {author} {\bibfnamefont
  {B.}~\bibnamefont {Keimer}},\ }\href {\doibase 10.1038/nphys720} {\bibfield
  {journal} {\bibinfo  {journal} {Nat. Phys.}\ }\textbf {\bibinfo {volume}
  {3}},\ \bibinfo {pages} {780} (\bibinfo {year} {2007})}\BibitemShut {NoStop}%
\bibitem [{\citenamefont {Norman}\ \emph {et~al.}(1995)\citenamefont {Norman},
  \citenamefont {Randeria}, \citenamefont {Ding},\ and\ \citenamefont
  {Campuzano}}]{Norman:1995dd}%
  \BibitemOpen
  \bibfield  {author} {\bibinfo {author} {\bibfnamefont {M.~R.}\ \bibnamefont
  {Norman}}, \bibinfo {author} {\bibfnamefont {M.}~\bibnamefont {Randeria}},
  \bibinfo {author} {\bibfnamefont {H.}~\bibnamefont {Ding}}, \ and\ \bibinfo
  {author} {\bibfnamefont {J.~C.}\ \bibnamefont {Campuzano}},\ }\href {\doibase
  10.1103/physrevb.52.615} {\bibfield  {journal} {\bibinfo  {journal} {Phys.
  Rev. B}\ }\textbf {\bibinfo {volume} {52}},\ \bibinfo {pages} {615} (\bibinfo
  {year} {1995})}\BibitemShut {NoStop}%
\bibitem [{\citenamefont {Banerjee}\ \emph
  {et~al.}(2011{\natexlab{a}})\citenamefont {Banerjee}, \citenamefont
  {Ramakrishnan},\ and\ \citenamefont {Dasgupta}}]{Banerjee:2011cu}%
  \BibitemOpen
  \bibfield  {author} {\bibinfo {author} {\bibfnamefont {S.}~\bibnamefont
  {Banerjee}}, \bibinfo {author} {\bibfnamefont {T.~V.}\ \bibnamefont
  {Ramakrishnan}}, \ and\ \bibinfo {author} {\bibfnamefont {C.}~\bibnamefont
  {Dasgupta}},\ }\href {\doibase 10.1103/physrevb.84.144525} {\bibfield
  {journal} {\bibinfo  {journal} {Phys. Rev. B}\ }\textbf {\bibinfo {volume}
  {84}},\ \bibinfo {pages} {144525} (\bibinfo {year}
  {2011}{\natexlab{a}})}\BibitemShut {NoStop}%
\bibitem [{\citenamefont {Banerjee}\ \emph
  {et~al.}(2011{\natexlab{b}})\citenamefont {Banerjee}, \citenamefont
  {Ramakrishnan},\ and\ \citenamefont {Dasgupta}}]{Banerjee:2011bz}%
  \BibitemOpen
  \bibfield  {author} {\bibinfo {author} {\bibfnamefont {S.}~\bibnamefont
  {Banerjee}}, \bibinfo {author} {\bibfnamefont {T.~V.}\ \bibnamefont
  {Ramakrishnan}}, \ and\ \bibinfo {author} {\bibfnamefont {C.}~\bibnamefont
  {Dasgupta}},\ }\href {\doibase 10.1103/PhysRevB.83.024510} {\bibfield
  {journal} {\bibinfo  {journal} {Phys. Rev. B}\ }\textbf {\bibinfo {volume}
  {83}},\ \bibinfo {pages} {024510} (\bibinfo {year}
  {2011}{\natexlab{b}})}\BibitemShut {NoStop}%
\bibitem [{\citenamefont {Norman}\ \emph {et~al.}(1998)\citenamefont {Norman},
  \citenamefont {Randeria}, \citenamefont {Ding},\ and\ \citenamefont
  {Campuzano}}]{Norman:1998va}%
  \BibitemOpen
  \bibfield  {author} {\bibinfo {author} {\bibfnamefont {M.~R.}\ \bibnamefont
  {Norman}}, \bibinfo {author} {\bibfnamefont {M.}~\bibnamefont {Randeria}},
  \bibinfo {author} {\bibfnamefont {H.}~\bibnamefont {Ding}}, \ and\ \bibinfo
  {author} {\bibfnamefont {J.~C.}\ \bibnamefont {Campuzano}},\ }\href {\doibase
  10.1103/PhysRevB.57.R11093} {\bibfield  {journal} {\bibinfo  {journal} {Phys.
  Rev. B}\ }\textbf {\bibinfo {volume} {57}},\ \bibinfo {pages} {R11093}
  (\bibinfo {year} {1998})}\BibitemShut {NoStop}%
\bibitem [{\citenamefont {Chien}\ \emph {et~al.}(2009)\citenamefont {Chien},
  \citenamefont {He}, \citenamefont {Chen},\ and\ \citenamefont
  {Levin}}]{Chien2009}%
  \BibitemOpen
  \bibfield  {author} {\bibinfo {author} {\bibfnamefont {C.-C.}\ \bibnamefont
  {Chien}}, \bibinfo {author} {\bibfnamefont {Y.}~\bibnamefont {He}}, \bibinfo
  {author} {\bibfnamefont {Q.}~\bibnamefont {Chen}}, \ and\ \bibinfo {author}
  {\bibfnamefont {K.}~\bibnamefont {Levin}},\ }\href@noop {} {\bibfield
  {journal} {\bibinfo  {journal} {Physical Review B}\ }\textbf {\bibinfo
  {volume} {79}},\ \bibinfo {pages} {214527} (\bibinfo {year}
  {2009})}\BibitemShut {NoStop}%
\bibitem [{\citenamefont {Campuzano}\ \emph {et~al.}(1998)\citenamefont
  {Campuzano}, \citenamefont {Norman}, \citenamefont {Ding}, \citenamefont
  {Randeria}, \citenamefont {Yokoya}, \citenamefont {Takeuchi}, \citenamefont
  {Takahashi}, \citenamefont {Mochiku}, \citenamefont {Kadowaki}, \citenamefont
  {Guptasarma},\ and\ \citenamefont {Hinks}}]{Campuzano98}%
  \BibitemOpen
  \bibfield  {author} {\bibinfo {author} {\bibfnamefont {J.~C.}\ \bibnamefont
  {Campuzano}}, \bibinfo {author} {\bibfnamefont {M.~R.}\ \bibnamefont
  {Norman}}, \bibinfo {author} {\bibfnamefont {H.}~\bibnamefont {Ding}},
  \bibinfo {author} {\bibfnamefont {M.}~\bibnamefont {Randeria}}, \bibinfo
  {author} {\bibfnamefont {T.}~\bibnamefont {Yokoya}}, \bibinfo {author}
  {\bibfnamefont {T.}~\bibnamefont {Takeuchi}}, \bibinfo {author}
  {\bibfnamefont {T.}~\bibnamefont {Takahashi}}, \bibinfo {author}
  {\bibfnamefont {T.}~\bibnamefont {Mochiku}}, \bibinfo {author} {\bibfnamefont
  {K.}~\bibnamefont {Kadowaki}}, \bibinfo {author} {\bibfnamefont
  {P.}~\bibnamefont {Guptasarma}}, \ and\ \bibinfo {author} {\bibfnamefont
  {D.~G.}\ \bibnamefont {Hinks}},\ }\href {\doibase 10.1038/32366} {\bibfield
  {journal} {\bibinfo  {journal} {Nature}\ }\textbf {\bibinfo {volume} {392}},\
  \bibinfo {pages} {157} (\bibinfo {year} {1998})}\BibitemShut {NoStop}%
\bibitem [{\citenamefont {Campuzano}\ \emph {et~al.}(1996)\citenamefont
  {Campuzano}, \citenamefont {Ding}, \citenamefont {Norman},\ and\
  \citenamefont {Randeira}}]{Campuzano:1996fb}%
  \BibitemOpen
  \bibfield  {author} {\bibinfo {author} {\bibfnamefont {J.~C.}\ \bibnamefont
  {Campuzano}}, \bibinfo {author} {\bibfnamefont {H.}~\bibnamefont {Ding}},
  \bibinfo {author} {\bibfnamefont {M.~R.}\ \bibnamefont {Norman}}, \ and\
  \bibinfo {author} {\bibfnamefont {M.}~\bibnamefont {Randeira}},\ }\href
  {\doibase 10.1103/physrevb.53.r14737} {\bibfield  {journal} {\bibinfo
  {journal} {Phys. Rev. B}\ }\textbf {\bibinfo {volume} {53}},\ \bibinfo
  {pages} {R14737} (\bibinfo {year} {1996})}\BibitemShut {NoStop}%
\bibitem [{\citenamefont {Benfatto}\ \emph {et~al.}(2000)\citenamefont
  {Benfatto}, \citenamefont {Caprara},\ and\ \citenamefont
  {Castro}}]{Benfatto00}%
  \BibitemOpen
  \bibfield  {author} {\bibinfo {author} {\bibfnamefont {L.}~\bibnamefont
  {Benfatto}}, \bibinfo {author} {\bibfnamefont {S.}~\bibnamefont {Caprara}}, \
  and\ \bibinfo {author} {\bibfnamefont {C.~D.}\ \bibnamefont {Castro}},\
  }\href {\doibase 10.1007/s100510070163} {\bibfield  {journal} {\bibinfo
  {journal} {Eur. Phys. J. B}\ }\textbf {\bibinfo {volume} {17}},\ \bibinfo
  {pages} {95} (\bibinfo {year} {2000})}\BibitemShut {NoStop}%
\bibitem [{\citenamefont {Del~Maestro}\ \emph {et~al.}(2006)\citenamefont
  {Del~Maestro}, \citenamefont {Rosenow},\ and\ \citenamefont
  {Sachdev}}]{DelMaestro06}%
  \BibitemOpen
  \bibfield  {author} {\bibinfo {author} {\bibfnamefont {A.}~\bibnamefont
  {Del~Maestro}}, \bibinfo {author} {\bibfnamefont {B.}~\bibnamefont
  {Rosenow}}, \ and\ \bibinfo {author} {\bibfnamefont {S.}~\bibnamefont
  {Sachdev}},\ }\href {\doibase 10.1103/PhysRevB.74.024520} {\bibfield
  {journal} {\bibinfo  {journal} {Phys. Rev. B}\ }\textbf {\bibinfo {volume}
  {74}},\ \bibinfo {pages} {024520} (\bibinfo {year} {2006})}\BibitemShut
  {NoStop}%
\bibitem [{\citenamefont {Loret}\ \emph {et~al.}(2019)\citenamefont {Loret},
  \citenamefont {Auvray}, \citenamefont {Gallais}, \citenamefont {Cazayous},
  \citenamefont {Forget}, \citenamefont {Colson}, \citenamefont {Julien},
  \citenamefont {Paul}, \citenamefont {Civelli},\ and\ \citenamefont
  {Sacuto}}]{Loret19}%
  \BibitemOpen
  \bibfield  {author} {\bibinfo {author} {\bibfnamefont {B.}~\bibnamefont
  {Loret}}, \bibinfo {author} {\bibfnamefont {N.}~\bibnamefont {Auvray}},
  \bibinfo {author} {\bibfnamefont {Y.}~\bibnamefont {Gallais}}, \bibinfo
  {author} {\bibfnamefont {M.}~\bibnamefont {Cazayous}}, \bibinfo {author}
  {\bibfnamefont {A.}~\bibnamefont {Forget}}, \bibinfo {author} {\bibfnamefont
  {D.}~\bibnamefont {Colson}}, \bibinfo {author} {\bibfnamefont {M.-H.}\
  \bibnamefont {Julien}}, \bibinfo {author} {\bibfnamefont {I.}~\bibnamefont
  {Paul}}, \bibinfo {author} {\bibfnamefont {M.}~\bibnamefont {Civelli}}, \
  and\ \bibinfo {author} {\bibfnamefont {A.}~\bibnamefont {Sacuto}},\ }\href
  {https://www.nature.com/articles/s41567-019-0509-5} {\bibfield  {journal}
  {\bibinfo  {journal} {Nature Physics}\ ,\ \bibinfo {pages} {1}} (\bibinfo
  {year} {2019})}\BibitemShut {NoStop}%
\bibitem [{\citenamefont {Chang}\ \emph {et~al.}(2012)\citenamefont {Chang},
  \citenamefont {Blackburn}, \citenamefont {Holmes}, \citenamefont
  {Christensen}, \citenamefont {Larsen}, \citenamefont {Mesot}, \citenamefont
  {Liang}, \citenamefont {Bonn}, \citenamefont {Hardy}, \citenamefont
  {Watenphul}, \citenamefont {Zimmermann}, \citenamefont {Forgan},\ and\
  \citenamefont {Hayden}}]{Chang12}%
  \BibitemOpen
  \bibfield  {author} {\bibinfo {author} {\bibfnamefont {J.}~\bibnamefont
  {Chang}}, \bibinfo {author} {\bibfnamefont {E.}~\bibnamefont {Blackburn}},
  \bibinfo {author} {\bibfnamefont {A.~T.}\ \bibnamefont {Holmes}}, \bibinfo
  {author} {\bibfnamefont {N.~B.}\ \bibnamefont {Christensen}}, \bibinfo
  {author} {\bibfnamefont {J.}~\bibnamefont {Larsen}}, \bibinfo {author}
  {\bibfnamefont {J.}~\bibnamefont {Mesot}}, \bibinfo {author} {\bibfnamefont
  {R.}~\bibnamefont {Liang}}, \bibinfo {author} {\bibfnamefont {D.~A.}\
  \bibnamefont {Bonn}}, \bibinfo {author} {\bibfnamefont {W.~N.}\ \bibnamefont
  {Hardy}}, \bibinfo {author} {\bibfnamefont {A.}~\bibnamefont {Watenphul}},
  \bibinfo {author} {\bibfnamefont {M.~v.}\ \bibnamefont {Zimmermann}},
  \bibinfo {author} {\bibfnamefont {E.~M.}\ \bibnamefont {Forgan}}, \ and\
  \bibinfo {author} {\bibfnamefont {S.~M.}\ \bibnamefont {Hayden}},\ }\href
  {\doibase 10.1038/nphys2456} {\bibfield  {journal} {\bibinfo  {journal} {Nat.
  Phys.}\ }\textbf {\bibinfo {volume} {8}},\ \bibinfo {pages} {871} (\bibinfo
  {year} {2012})}\BibitemShut {NoStop}%
\bibitem [{\citenamefont {Blanco-Canosa}\ \emph {et~al.}(2013)\citenamefont
  {Blanco-Canosa}, \citenamefont {Frano}, \citenamefont {Loew}, \citenamefont
  {Lu}, \citenamefont {Porras}, \citenamefont {Ghiringhelli}, \citenamefont
  {Minola}, \citenamefont {Mazzoli}, \citenamefont {Braicovich}, \citenamefont
  {Schierle}, \citenamefont {Weschke}, \citenamefont {Le~Tacon},\ and\
  \citenamefont {Keimer}}]{Blanco-Canosa13}%
  \BibitemOpen
  \bibfield  {author} {\bibinfo {author} {\bibfnamefont {S.}~\bibnamefont
  {Blanco-Canosa}}, \bibinfo {author} {\bibfnamefont {A.}~\bibnamefont
  {Frano}}, \bibinfo {author} {\bibfnamefont {T.}~\bibnamefont {Loew}},
  \bibinfo {author} {\bibfnamefont {Y.}~\bibnamefont {Lu}}, \bibinfo {author}
  {\bibfnamefont {J.}~\bibnamefont {Porras}}, \bibinfo {author} {\bibfnamefont
  {G.}~\bibnamefont {Ghiringhelli}}, \bibinfo {author} {\bibfnamefont
  {M.}~\bibnamefont {Minola}}, \bibinfo {author} {\bibfnamefont
  {C.}~\bibnamefont {Mazzoli}}, \bibinfo {author} {\bibfnamefont
  {L.}~\bibnamefont {Braicovich}}, \bibinfo {author} {\bibfnamefont
  {E.}~\bibnamefont {Schierle}}, \bibinfo {author} {\bibfnamefont
  {E.}~\bibnamefont {Weschke}}, \bibinfo {author} {\bibfnamefont
  {M.}~\bibnamefont {Le~Tacon}}, \ and\ \bibinfo {author} {\bibfnamefont
  {B.}~\bibnamefont {Keimer}},\ }\href {\doibase
  10.1103/PhysRevLett.110.187001} {\bibfield  {journal} {\bibinfo  {journal}
  {Phys. Rev. Lett.}\ }\textbf {\bibinfo {volume} {110}},\ \bibinfo {pages}
  {187001} (\bibinfo {year} {2013})}\BibitemShut {NoStop}%
\bibitem [{\citenamefont {Blackburn}\ \emph {et~al.}(2013)\citenamefont
  {Blackburn}, \citenamefont {Chang}, \citenamefont {H\"ucker}, \citenamefont
  {Holmes}, \citenamefont {Christensen}, \citenamefont {Liang}, \citenamefont
  {Bonn}, \citenamefont {Hardy}, \citenamefont {R\"utt}, \citenamefont
  {Gutowski}, \citenamefont {Zimmermann}, \citenamefont {Forgan},\ and\
  \citenamefont {Hayden}}]{Blackburn13a}%
  \BibitemOpen
  \bibfield  {author} {\bibinfo {author} {\bibfnamefont {E.}~\bibnamefont
  {Blackburn}}, \bibinfo {author} {\bibfnamefont {J.}~\bibnamefont {Chang}},
  \bibinfo {author} {\bibfnamefont {M.}~\bibnamefont {H\"ucker}}, \bibinfo
  {author} {\bibfnamefont {A.~T.}\ \bibnamefont {Holmes}}, \bibinfo {author}
  {\bibfnamefont {N.~B.}\ \bibnamefont {Christensen}}, \bibinfo {author}
  {\bibfnamefont {R.}~\bibnamefont {Liang}}, \bibinfo {author} {\bibfnamefont
  {D.~A.}\ \bibnamefont {Bonn}}, \bibinfo {author} {\bibfnamefont {W.~N.}\
  \bibnamefont {Hardy}}, \bibinfo {author} {\bibfnamefont {U.}~\bibnamefont
  {R\"utt}}, \bibinfo {author} {\bibfnamefont {O.}~\bibnamefont {Gutowski}},
  \bibinfo {author} {\bibfnamefont {M.~v.}\ \bibnamefont {Zimmermann}},
  \bibinfo {author} {\bibfnamefont {E.~M.}\ \bibnamefont {Forgan}}, \ and\
  \bibinfo {author} {\bibfnamefont {S.~M.}\ \bibnamefont {Hayden}},\ }\href
  {\doibase 10.1103/PhysRevLett.110.137004} {\bibfield  {journal} {\bibinfo
  {journal} {Phys. Rev. Lett.}\ }\textbf {\bibinfo {volume} {110}},\ \bibinfo
  {pages} {137004} (\bibinfo {year} {2013})}\BibitemShut {NoStop}%
\bibitem [{\citenamefont {Ghiringhelli}\ \emph {et~al.}(2012)\citenamefont
  {Ghiringhelli}, \citenamefont {Le~Tacon}, \citenamefont {Minola},
  \citenamefont {Blanco-Canosa}, \citenamefont {Mazzoli}, \citenamefont
  {Brookes}, \citenamefont {De~Luca}, \citenamefont {Frano}, \citenamefont
  {Hawthorn}, \citenamefont {He}, \citenamefont {Loew}, \citenamefont {Sala},
  \citenamefont {Peets}, \citenamefont {Salluzzo}, \citenamefont {Schierle},
  \citenamefont {Sutarto}, \citenamefont {Sawatzky}, \citenamefont {Weschke},
  \citenamefont {Keimer},\ and\ \citenamefont {Braicovich}}]{Ghiringhelli12}%
  \BibitemOpen
  \bibfield  {author} {\bibinfo {author} {\bibfnamefont {G.}~\bibnamefont
  {Ghiringhelli}}, \bibinfo {author} {\bibfnamefont {M.}~\bibnamefont
  {Le~Tacon}}, \bibinfo {author} {\bibfnamefont {M.}~\bibnamefont {Minola}},
  \bibinfo {author} {\bibfnamefont {S.}~\bibnamefont {Blanco-Canosa}}, \bibinfo
  {author} {\bibfnamefont {C.}~\bibnamefont {Mazzoli}}, \bibinfo {author}
  {\bibfnamefont {N.~B.}\ \bibnamefont {Brookes}}, \bibinfo {author}
  {\bibfnamefont {G.~M.}\ \bibnamefont {De~Luca}}, \bibinfo {author}
  {\bibfnamefont {A.}~\bibnamefont {Frano}}, \bibinfo {author} {\bibfnamefont
  {D.~G.}\ \bibnamefont {Hawthorn}}, \bibinfo {author} {\bibfnamefont
  {F.}~\bibnamefont {He}}, \bibinfo {author} {\bibfnamefont {T.}~\bibnamefont
  {Loew}}, \bibinfo {author} {\bibfnamefont {M.~M.}\ \bibnamefont {Sala}},
  \bibinfo {author} {\bibfnamefont {D.~C.}\ \bibnamefont {Peets}}, \bibinfo
  {author} {\bibfnamefont {M.}~\bibnamefont {Salluzzo}}, \bibinfo {author}
  {\bibfnamefont {E.}~\bibnamefont {Schierle}}, \bibinfo {author}
  {\bibfnamefont {R.}~\bibnamefont {Sutarto}}, \bibinfo {author} {\bibfnamefont
  {G.~A.}\ \bibnamefont {Sawatzky}}, \bibinfo {author} {\bibfnamefont
  {E.}~\bibnamefont {Weschke}}, \bibinfo {author} {\bibfnamefont
  {B.}~\bibnamefont {Keimer}}, \ and\ \bibinfo {author} {\bibfnamefont
  {L.}~\bibnamefont {Braicovich}},\ }\href {\doibase 10.1126/science.1223532}
  {\bibfield  {journal} {\bibinfo  {journal} {Science}\ }\textbf {\bibinfo
  {volume} {337}},\ \bibinfo {pages} {821} (\bibinfo {year}
  {2012})}\BibitemShut {NoStop}%
\bibitem [{\citenamefont {Gerber}\ \emph {et~al.}(2015)\citenamefont {Gerber},
  \citenamefont {Jang}, \citenamefont {Nojiri}, \citenamefont {Matsuzawa},
  \citenamefont {Yasumura}, \citenamefont {Bonn}, \citenamefont {Liang},
  \citenamefont {Hardy}, \citenamefont {Islam}, \citenamefont {Mehta},
  \citenamefont {Song}, \citenamefont {Sikorski}, \citenamefont {Stefanescu},
  \citenamefont {Feng}, \citenamefont {Kivelson}, \citenamefont {Devereaux},
  \citenamefont {Shen}, \citenamefont {Kao}, \citenamefont {Lee}, \citenamefont
  {Zhu},\ and\ \citenamefont {Lee}}]{Gerber:2015gx}%
  \BibitemOpen
  \bibfield  {author} {\bibinfo {author} {\bibfnamefont {S.}~\bibnamefont
  {Gerber}}, \bibinfo {author} {\bibfnamefont {H.}~\bibnamefont {Jang}},
  \bibinfo {author} {\bibfnamefont {H.}~\bibnamefont {Nojiri}}, \bibinfo
  {author} {\bibfnamefont {S.}~\bibnamefont {Matsuzawa}}, \bibinfo {author}
  {\bibfnamefont {H.}~\bibnamefont {Yasumura}}, \bibinfo {author}
  {\bibfnamefont {D.~A.}\ \bibnamefont {Bonn}}, \bibinfo {author}
  {\bibfnamefont {R.}~\bibnamefont {Liang}}, \bibinfo {author} {\bibfnamefont
  {W.~N.}\ \bibnamefont {Hardy}}, \bibinfo {author} {\bibfnamefont
  {Z.}~\bibnamefont {Islam}}, \bibinfo {author} {\bibfnamefont
  {A.}~\bibnamefont {Mehta}}, \bibinfo {author} {\bibfnamefont
  {S.}~\bibnamefont {Song}}, \bibinfo {author} {\bibfnamefont {M.}~\bibnamefont
  {Sikorski}}, \bibinfo {author} {\bibfnamefont {D.}~\bibnamefont
  {Stefanescu}}, \bibinfo {author} {\bibfnamefont {Y.}~\bibnamefont {Feng}},
  \bibinfo {author} {\bibfnamefont {S.~A.}\ \bibnamefont {Kivelson}}, \bibinfo
  {author} {\bibfnamefont {T.~P.}\ \bibnamefont {Devereaux}}, \bibinfo {author}
  {\bibfnamefont {Z.-X.}\ \bibnamefont {Shen}}, \bibinfo {author}
  {\bibfnamefont {C.~C.}\ \bibnamefont {Kao}}, \bibinfo {author} {\bibfnamefont
  {W.~S.}\ \bibnamefont {Lee}}, \bibinfo {author} {\bibfnamefont
  {D.}~\bibnamefont {Zhu}}, \ and\ \bibinfo {author} {\bibfnamefont {J.~S.}\
  \bibnamefont {Lee}},\ }\href {\doibase 10.1126/science.aac6257} {\bibfield
  {journal} {\bibinfo  {journal} {Science}\ }\textbf {\bibinfo {volume}
  {350}},\ \bibinfo {pages} {949} (\bibinfo {year} {2015})}\BibitemShut
  {NoStop}%
\bibitem [{\citenamefont {Chang}\ \emph {et~al.}(2016)\citenamefont {Chang},
  \citenamefont {Blackburn}, \citenamefont {Ivashko}, \citenamefont {Holmes},
  \citenamefont {Christensen}, \citenamefont {Hucker}, \citenamefont {Liang},
  \citenamefont {Bonn}, \citenamefont {Hardy}, \citenamefont {Rutt},
  \citenamefont {Zimmermann}, \citenamefont {Forgan},\ and\ \citenamefont
  {M.}}]{Chang16}%
  \BibitemOpen
  \bibfield  {author} {\bibinfo {author} {\bibfnamefont {J.}~\bibnamefont
  {Chang}}, \bibinfo {author} {\bibfnamefont {E.}~\bibnamefont {Blackburn}},
  \bibinfo {author} {\bibfnamefont {O.}~\bibnamefont {Ivashko}}, \bibinfo
  {author} {\bibfnamefont {A.~T.}\ \bibnamefont {Holmes}}, \bibinfo {author}
  {\bibfnamefont {N.~B.}\ \bibnamefont {Christensen}}, \bibinfo {author}
  {\bibfnamefont {M.}~\bibnamefont {Hucker}}, \bibinfo {author} {\bibfnamefont
  {R.}~\bibnamefont {Liang}}, \bibinfo {author} {\bibfnamefont {D.~A.}\
  \bibnamefont {Bonn}}, \bibinfo {author} {\bibfnamefont {W.~N.}\ \bibnamefont
  {Hardy}}, \bibinfo {author} {\bibfnamefont {U.}~\bibnamefont {Rutt}},
  \bibinfo {author} {\bibfnamefont {M.~v.}\ \bibnamefont {Zimmermann}},
  \bibinfo {author} {\bibfnamefont {E.~M.}\ \bibnamefont {Forgan}}, \ and\
  \bibinfo {author} {\bibfnamefont {H.~S.}\ \bibnamefont {M.}},\ }\href
  {http://http://www.nature.com/articles/ncomms11494} {\bibfield  {journal}
  {\bibinfo  {journal} {Nat. Commun.}\ }\textbf {\bibinfo {volume} {7}},\
  \bibinfo {pages} {11494} (\bibinfo {year} {2016})}\BibitemShut {NoStop}%
\bibitem [{\citenamefont {Wu}\ \emph {et~al.}(2011)\citenamefont {Wu},
  \citenamefont {Mayaffre}, \citenamefont {Kr{\"a}mer}, \citenamefont
  {Horvatic}, \citenamefont {Berthier}, \citenamefont {Hardy}, \citenamefont
  {Liang}, \citenamefont {Bonn},\ and\ \citenamefont {Julien}}]{Wu11}%
  \BibitemOpen
  \bibfield  {author} {\bibinfo {author} {\bibfnamefont {T.}~\bibnamefont
  {Wu}}, \bibinfo {author} {\bibfnamefont {H.}~\bibnamefont {Mayaffre}},
  \bibinfo {author} {\bibfnamefont {S.}~\bibnamefont {Kr{\"a}mer}}, \bibinfo
  {author} {\bibfnamefont {M.}~\bibnamefont {Horvatic}}, \bibinfo {author}
  {\bibfnamefont {C.}~\bibnamefont {Berthier}}, \bibinfo {author}
  {\bibfnamefont {W.~N.}\ \bibnamefont {Hardy}}, \bibinfo {author}
  {\bibfnamefont {R.}~\bibnamefont {Liang}}, \bibinfo {author} {\bibfnamefont
  {D.~A.}\ \bibnamefont {Bonn}}, \ and\ \bibinfo {author} {\bibfnamefont
  {M.-H.}\ \bibnamefont {Julien}},\ }\href {\doibase 10.1038/nature10345}
  {\bibfield  {journal} {\bibinfo  {journal} {Nature}\ }\textbf {\bibinfo
  {volume} {477}},\ \bibinfo {pages} {191} (\bibinfo {year}
  {2011})}\BibitemShut {NoStop}%
\bibitem [{\citenamefont {Wu}\ \emph {et~al.}(2013)\citenamefont {Wu},
  \citenamefont {Mayaffre}, \citenamefont {Kr{\"a}mer}, \citenamefont
  {Horvati{\'c}}, \citenamefont {Berthier}, \citenamefont {Kuhns},
  \citenamefont {Reyes}, \citenamefont {Liang}, \citenamefont {Hardy},
  \citenamefont {Bonn},\ and\ \citenamefont {Julien}}]{Wu13a}%
  \BibitemOpen
  \bibfield  {author} {\bibinfo {author} {\bibfnamefont {T.}~\bibnamefont
  {Wu}}, \bibinfo {author} {\bibfnamefont {H.}~\bibnamefont {Mayaffre}},
  \bibinfo {author} {\bibfnamefont {S.}~\bibnamefont {Kr{\"a}mer}}, \bibinfo
  {author} {\bibfnamefont {M.}~\bibnamefont {Horvati{\'c}}}, \bibinfo {author}
  {\bibfnamefont {C.}~\bibnamefont {Berthier}}, \bibinfo {author}
  {\bibfnamefont {P.~L.}\ \bibnamefont {Kuhns}}, \bibinfo {author}
  {\bibfnamefont {A.~P.}\ \bibnamefont {Reyes}}, \bibinfo {author}
  {\bibfnamefont {R.}~\bibnamefont {Liang}}, \bibinfo {author} {\bibfnamefont
  {W.~N.}\ \bibnamefont {Hardy}}, \bibinfo {author} {\bibfnamefont {D.~A.}\
  \bibnamefont {Bonn}}, \ and\ \bibinfo {author} {\bibfnamefont {M.-H.}\
  \bibnamefont {Julien}},\ }\href {\doibase 10.1038/ncomms3113} {\bibfield
  {journal} {\bibinfo  {journal} {Nat. Commun.}\ }\textbf {\bibinfo {volume}
  {4}},\ \bibinfo {pages} {2113} (\bibinfo {year} {2013})}\BibitemShut
  {NoStop}%
\bibitem [{\citenamefont {{Wu}}\ \emph {et~al.}(2015)\citenamefont {{Wu}},
  \citenamefont {{Mayaffre}}, \citenamefont {{Kr{\"a}mer}}, \citenamefont
  {{Horvati{\'c}}}, \citenamefont {{Berthier}}, \citenamefont {{Hardy}},
  \citenamefont {{Liang}}, \citenamefont {{Bonn}},\ and\ \citenamefont
  {{Julien}}}]{Wu:2015bt}%
  \BibitemOpen
  \bibfield  {author} {\bibinfo {author} {\bibfnamefont {T.}~\bibnamefont
  {{Wu}}}, \bibinfo {author} {\bibfnamefont {H.}~\bibnamefont {{Mayaffre}}},
  \bibinfo {author} {\bibfnamefont {S.}~\bibnamefont {{Kr{\"a}mer}}}, \bibinfo
  {author} {\bibfnamefont {M.}~\bibnamefont {{Horvati{\'c}}}}, \bibinfo
  {author} {\bibfnamefont {C.}~\bibnamefont {{Berthier}}}, \bibinfo {author}
  {\bibfnamefont {W.~N.}\ \bibnamefont {{Hardy}}}, \bibinfo {author}
  {\bibfnamefont {R.}~\bibnamefont {{Liang}}}, \bibinfo {author} {\bibfnamefont
  {D.~A.}\ \bibnamefont {{Bonn}}}, \ and\ \bibinfo {author} {\bibfnamefont
  {M.-H.}\ \bibnamefont {{Julien}}},\ }\href {\doibase 10.1038/ncomms7438}
  {\bibfield  {journal} {\bibinfo  {journal} {Nature Communications}\ }\textbf
  {\bibinfo {volume} {6}},\ \bibinfo {eid} {6438} (\bibinfo {year}
  {2015})}\BibitemShut {NoStop}%
\bibitem [{\citenamefont {Julien}(2015)}]{Julien15}%
  \BibitemOpen
  \bibfield  {author} {\bibinfo {author} {\bibfnamefont {M.-H.}\ \bibnamefont
  {Julien}},\ }\href {\doibase 10.1126/science.aad3279} {\bibfield  {journal}
  {\bibinfo  {journal} {Science}\ }\textbf {\bibinfo {volume} {350}},\ \bibinfo
  {pages} {914} (\bibinfo {year} {2015})}\BibitemShut {NoStop}%
\bibitem [{\citenamefont {Koley}\ \emph {et~al.}(2020)\citenamefont {Koley},
  \citenamefont {Mohanta},\ and\ \citenamefont {Taraphder}}]{Koley20}%
  \BibitemOpen
  \bibfield  {author} {\bibinfo {author} {\bibfnamefont {S.}~\bibnamefont
  {Koley}}, \bibinfo {author} {\bibfnamefont {N.}~\bibnamefont {Mohanta}}, \
  and\ \bibinfo {author} {\bibfnamefont {A.}~\bibnamefont {Taraphder}},\ }\href
  {\doibase 10.1140/epjb/e2020-100522-5} {\bibfield  {journal} {\bibinfo
  {journal} {The European Physical Journal B}\ }\textbf {\bibinfo {volume}
  {93}} (\bibinfo {year} {2020}),\ 10.1140/epjb/e2020-100522-5}\BibitemShut
  {NoStop}%
\bibitem [{\citenamefont {Kawasaki}\ \emph {et~al.}(2005)\citenamefont
  {Kawasaki}, \citenamefont {Yashima}, \citenamefont {Mito}, \citenamefont
  {Kawasaki}, \citenamefont {Zheng}, \citenamefont {Kitaoka}, \citenamefont
  {Aoki}, \citenamefont {Haga},\ and\ \citenamefont {Ōnuki}}]{Kawasaki05}%
  \BibitemOpen
  \bibfield  {author} {\bibinfo {author} {\bibfnamefont {S.}~\bibnamefont
  {Kawasaki}}, \bibinfo {author} {\bibfnamefont {M.}~\bibnamefont {Yashima}},
  \bibinfo {author} {\bibfnamefont {T.}~\bibnamefont {Mito}}, \bibinfo {author}
  {\bibfnamefont {Y.}~\bibnamefont {Kawasaki}}, \bibinfo {author}
  {\bibfnamefont {G.-Q.}\ \bibnamefont {Zheng}}, \bibinfo {author}
  {\bibfnamefont {Y.}~\bibnamefont {Kitaoka}}, \bibinfo {author} {\bibfnamefont
  {D.}~\bibnamefont {Aoki}}, \bibinfo {author} {\bibfnamefont {Y.}~\bibnamefont
  {Haga}}, \ and\ \bibinfo {author} {\bibfnamefont {Y.}~\bibnamefont
  {Ōnuki}},\ }\href {\doibase 10.1088/0953-8984/17/11/019} {\bibfield
  {journal} {\bibinfo  {journal} {Journal of Physics: Condensed Matter}\
  }\textbf {\bibinfo {volume} {17}} (\bibinfo {year} {2005}),\
  10.1088/0953-8984/17/11/019}\BibitemShut {NoStop}%
\bibitem [{\citenamefont {Borisenko}\ \emph {et~al.}(2009)\citenamefont
  {Borisenko}, \citenamefont {Kordyuk}, \citenamefont {Zabolotnyy},
  \citenamefont {Inosov}, \citenamefont {Evtushinsky}, \citenamefont
  {Büchner}, \citenamefont {Yaresko}, \citenamefont {Varykhalov},
  \citenamefont {Follath}, \citenamefont {Eberhardt},\ and\ \citenamefont
  {et~al.}}]{Borisenko09}%
  \BibitemOpen
  \bibfield  {author} {\bibinfo {author} {\bibfnamefont {S.~V.}\ \bibnamefont
  {Borisenko}}, \bibinfo {author} {\bibfnamefont {A.~A.}\ \bibnamefont
  {Kordyuk}}, \bibinfo {author} {\bibfnamefont {V.~B.}\ \bibnamefont
  {Zabolotnyy}}, \bibinfo {author} {\bibfnamefont {D.~S.}\ \bibnamefont
  {Inosov}}, \bibinfo {author} {\bibfnamefont {D.}~\bibnamefont {Evtushinsky}},
  \bibinfo {author} {\bibfnamefont {B.}~\bibnamefont {Büchner}}, \bibinfo
  {author} {\bibfnamefont {A.~N.}\ \bibnamefont {Yaresko}}, \bibinfo {author}
  {\bibfnamefont {A.}~\bibnamefont {Varykhalov}}, \bibinfo {author}
  {\bibfnamefont {R.}~\bibnamefont {Follath}}, \bibinfo {author} {\bibfnamefont
  {W.}~\bibnamefont {Eberhardt}}, \ and\ \bibinfo {author} {\bibnamefont
  {et~al.}},\ }\href {\doibase 10.1103/physrevlett.102.166402} {\bibfield
  {journal} {\bibinfo  {journal} {Physical Review Letters}\ }\textbf {\bibinfo
  {volume} {102}} (\bibinfo {year} {2009}),\
  10.1103/physrevlett.102.166402}\BibitemShut {NoStop}%
\bibitem [{\citenamefont {Chatterjee}\ \emph {et~al.}(2015)\citenamefont
  {Chatterjee}, \citenamefont {Zhao}, \citenamefont {Iavarone}, \citenamefont
  {Capua}, \citenamefont {Castellan}, \citenamefont {Karapetrov}, \citenamefont
  {Malliakas}, \citenamefont {Kanatzidis}, \citenamefont {Claus}, \citenamefont
  {Ruff},\ and\ \citenamefont {et~al.}}]{Chatterjee15}%
  \BibitemOpen
  \bibfield  {author} {\bibinfo {author} {\bibfnamefont {U.}~\bibnamefont
  {Chatterjee}}, \bibinfo {author} {\bibfnamefont {J.}~\bibnamefont {Zhao}},
  \bibinfo {author} {\bibfnamefont {M.}~\bibnamefont {Iavarone}}, \bibinfo
  {author} {\bibfnamefont {R.~D.}\ \bibnamefont {Capua}}, \bibinfo {author}
  {\bibfnamefont {J.~P.}\ \bibnamefont {Castellan}}, \bibinfo {author}
  {\bibfnamefont {G.}~\bibnamefont {Karapetrov}}, \bibinfo {author}
  {\bibfnamefont {C.~D.}\ \bibnamefont {Malliakas}}, \bibinfo {author}
  {\bibfnamefont {M.~G.}\ \bibnamefont {Kanatzidis}}, \bibinfo {author}
  {\bibfnamefont {H.}~\bibnamefont {Claus}}, \bibinfo {author} {\bibfnamefont
  {J.~P.~C.}\ \bibnamefont {Ruff}}, \ and\ \bibinfo {author} {\bibnamefont
  {et~al.}},\ }\href {\doibase 10.1038/ncomms7313} {\bibfield  {journal}
  {\bibinfo  {journal} {Nature Communications}\ }\textbf {\bibinfo {volume}
  {6}} (\bibinfo {year} {2015}),\ 10.1038/ncomms7313}\BibitemShut {NoStop}%
\end{thebibliography}%


\begin{thebibliography}{7}%
\makeatletter
\providecommand \@ifxundefined [1]{%
 \@ifx{#1\undefined}
}%
\providecommand \@ifnum [1]{%
 \ifnum #1\expandafter \@firstoftwo
 \else \expandafter \@secondoftwo
 \fi
}%
\providecommand \@ifx [1]{%
 \ifx #1\expandafter \@firstoftwo
 \else \expandafter \@secondoftwo
 \fi
}%
\providecommand \natexlab [1]{#1}%
\providecommand \enquote  [1]{``#1''}%
\providecommand \bibnamefont  [1]{#1}%
\providecommand \bibfnamefont [1]{#1}%
\providecommand \citenamefont [1]{#1}%
\providecommand \href@noop [0]{\@secondoftwo}%
\providecommand \href [0]{\begingroup \@sanitize@url \@href}%
\providecommand \@href[1]{\@@startlink{#1}\@@href}%
\providecommand \@@href[1]{\endgroup#1\@@endlink}%
\providecommand \@sanitize@url [0]{\catcode `\\12\catcode `\$12\catcode
  `\&12\catcode `\#12\catcode `\^12\catcode `\_12\catcode `\%12\relax}%
\providecommand \@@startlink[1]{}%
\providecommand \@@endlink[0]{}%
\providecommand \url  [0]{\begingroup\@sanitize@url \@url }%
\providecommand \@url [1]{\endgroup\@href {#1}{\urlprefix }}%
\providecommand \urlprefix  [0]{URL }%
\providecommand \Eprint [0]{\href }%
\providecommand \doibase [0]{http://dx.doi.org/}%
\providecommand \selectlanguage [0]{\@gobble}%
\providecommand \bibinfo  [0]{\@secondoftwo}%
\providecommand \bibfield  [0]{\@secondoftwo}%
\providecommand \translation [1]{[#1]}%
\providecommand \BibitemOpen [0]{}%
\providecommand \bibitemStop [0]{}%
\providecommand \bibitemNoStop [0]{.\EOS\space}%
\providecommand \EOS [0]{\spacefactor3000\relax}%
\providecommand \BibitemShut  [1]{\csname bibitem#1\endcsname}%
\let\auto@bib@innerbib\@empty
\bibitem [{\citenamefont {{Chakraborty}}\ \emph {et~al.}(2019)\citenamefont
  {{Chakraborty}}, \citenamefont {{Grandadam}}, \citenamefont {{Hamidian}},
  \citenamefont {{Davis}}, \citenamefont {{Sidis}},\ and\ \citenamefont
  {{P{\'e}pin}}}]{Chakraborty19}%
  \BibitemOpen
  \bibfield  {author} {\bibinfo {author} {\bibfnamefont {D.}~\bibnamefont
  {{Chakraborty}}}, \bibinfo {author} {\bibfnamefont {M.}~\bibnamefont
  {{Grandadam}}}, \bibinfo {author} {\bibfnamefont {M.~H.}\ \bibnamefont
  {{Hamidian}}}, \bibinfo {author} {\bibfnamefont {J.~C.~S.}\ \bibnamefont
  {{Davis}}}, \bibinfo {author} {\bibfnamefont {Y.}~\bibnamefont {{Sidis}}}, \
  and\ \bibinfo {author} {\bibfnamefont {C.}~\bibnamefont {{P{\'e}pin}}},\
  }\href@noop {} {\bibfield  {journal} {\bibinfo  {journal} {arXiv e-prints}\
  ,\ \bibinfo {eid} {arXiv:1906.01633}} (\bibinfo {year} {2019})},\ \Eprint
  {http://arxiv.org/abs/1906.01633} {arXiv:1906.01633 [cond-mat.supr-con]}
  \BibitemShut {NoStop}%
\bibitem [{\citenamefont {{Grandadam}}\ \emph {et~al.}(2019)\citenamefont
  {{Grandadam}}, \citenamefont {{Chakraborty}},\ and\ \citenamefont
  {{P{\'e}pin}}}]{Grandadam19}%
  \BibitemOpen
  \bibfield  {author} {\bibinfo {author} {\bibfnamefont {M.}~\bibnamefont
  {{Grandadam}}}, \bibinfo {author} {\bibfnamefont {D.}~\bibnamefont
  {{Chakraborty}}}, \ and\ \bibinfo {author} {\bibfnamefont {C.}~\bibnamefont
  {{P{\'e}pin}}},\ }\href@noop {} {\bibfield  {journal} {\bibinfo  {journal}
  {arXiv e-prints}\ ,\ \bibinfo {eid} {arXiv:1909.06657}} (\bibinfo {year}
  {2019})},\ \Eprint {http://arxiv.org/abs/1909.06657} {arXiv:1909.06657
  [cond-mat.supr-con]} \BibitemShut {NoStop}%
\bibitem [{\citenamefont {He}\ \emph {et~al.}(2011)\citenamefont {He},
  \citenamefont {Hashimoto}, \citenamefont {Karapetyan}, \citenamefont
  {Koralek}, \citenamefont {Hinton}, \citenamefont {Testaud}, \citenamefont
  {Nathan}, \citenamefont {Yoshida}, \citenamefont {Yao}, \citenamefont
  {Tanaka}, \citenamefont {Meevasana}, \citenamefont {Moore}, \citenamefont
  {Lu}, \citenamefont {Mo}, \citenamefont {Ishikado}, \citenamefont {Eisaki},
  \citenamefont {Hussain}, \citenamefont {Devereaux}, \citenamefont {Kivelson},
  \citenamefont {Orenstein}, \citenamefont {Kapitulnik},\ and\ \citenamefont
  {Shen}}]{He11}%
  \BibitemOpen
  \bibfield  {author} {\bibinfo {author} {\bibfnamefont {R.-H.}\ \bibnamefont
  {He}}, \bibinfo {author} {\bibfnamefont {M.}~\bibnamefont {Hashimoto}},
  \bibinfo {author} {\bibfnamefont {H.}~\bibnamefont {Karapetyan}}, \bibinfo
  {author} {\bibfnamefont {J.~D.}\ \bibnamefont {Koralek}}, \bibinfo {author}
  {\bibfnamefont {J.~P.}\ \bibnamefont {Hinton}}, \bibinfo {author}
  {\bibfnamefont {J.~P.}\ \bibnamefont {Testaud}}, \bibinfo {author}
  {\bibfnamefont {V.}~\bibnamefont {Nathan}}, \bibinfo {author} {\bibfnamefont
  {Y.}~\bibnamefont {Yoshida}}, \bibinfo {author} {\bibfnamefont
  {H.}~\bibnamefont {Yao}}, \bibinfo {author} {\bibfnamefont {K.}~\bibnamefont
  {Tanaka}}, \bibinfo {author} {\bibfnamefont {W.}~\bibnamefont {Meevasana}},
  \bibinfo {author} {\bibfnamefont {R.~G.}\ \bibnamefont {Moore}}, \bibinfo
  {author} {\bibfnamefont {D.~H.}\ \bibnamefont {Lu}}, \bibinfo {author}
  {\bibfnamefont {S.-K.}\ \bibnamefont {Mo}}, \bibinfo {author} {\bibfnamefont
  {M.}~\bibnamefont {Ishikado}}, \bibinfo {author} {\bibfnamefont
  {H.}~\bibnamefont {Eisaki}}, \bibinfo {author} {\bibfnamefont
  {Z.}~\bibnamefont {Hussain}}, \bibinfo {author} {\bibfnamefont {T.~P.}\
  \bibnamefont {Devereaux}}, \bibinfo {author} {\bibfnamefont {S.~A.}\
  \bibnamefont {Kivelson}}, \bibinfo {author} {\bibfnamefont {J.}~\bibnamefont
  {Orenstein}}, \bibinfo {author} {\bibfnamefont {A.}~\bibnamefont
  {Kapitulnik}}, \ and\ \bibinfo {author} {\bibfnamefont {Z.-X.}\ \bibnamefont
  {Shen}},\ }\href {\doibase 10.1126/science.1198415} {\bibfield  {journal}
  {\bibinfo  {journal} {Science}\ }\textbf {\bibinfo {volume} {331}},\ \bibinfo
  {pages} {1579} (\bibinfo {year} {2011})}\BibitemShut {NoStop}%
\bibitem [{\citenamefont {Norman}\ \emph {et~al.}(1998)\citenamefont {Norman},
  \citenamefont {Randeria}, \citenamefont {Ding},\ and\ \citenamefont
  {Campuzano}}]{Norman:1998va}%
  \BibitemOpen
  \bibfield  {author} {\bibinfo {author} {\bibfnamefont {M.~R.}\ \bibnamefont
  {Norman}}, \bibinfo {author} {\bibfnamefont {M.}~\bibnamefont {Randeria}},
  \bibinfo {author} {\bibfnamefont {H.}~\bibnamefont {Ding}}, \ and\ \bibinfo
  {author} {\bibfnamefont {J.~C.}\ \bibnamefont {Campuzano}},\ }\href {\doibase
  10.1103/PhysRevB.57.R11093} {\bibfield  {journal} {\bibinfo  {journal} {Phys.
  Rev. B}\ }\textbf {\bibinfo {volume} {57}},\ \bibinfo {pages} {R11093}
  (\bibinfo {year} {1998})}\BibitemShut {NoStop}%
\bibitem [{\citenamefont {Chien}\ \emph {et~al.}(2009)\citenamefont {Chien},
  \citenamefont {He}, \citenamefont {Chen},\ and\ \citenamefont
  {Levin}}]{Chien2009}%
  \BibitemOpen
  \bibfield  {author} {\bibinfo {author} {\bibfnamefont {C.-C.}\ \bibnamefont
  {Chien}}, \bibinfo {author} {\bibfnamefont {Y.}~\bibnamefont {He}}, \bibinfo
  {author} {\bibfnamefont {Q.}~\bibnamefont {Chen}}, \ and\ \bibinfo {author}
  {\bibfnamefont {K.}~\bibnamefont {Levin}},\ }\href@noop {} {\bibfield
  {journal} {\bibinfo  {journal} {Physical Review B}\ }\textbf {\bibinfo
  {volume} {79}},\ \bibinfo {pages} {214527} (\bibinfo {year}
  {2009})}\BibitemShut {NoStop}%
\bibitem [{\citenamefont {Campuzano}\ \emph {et~al.}(1998)\citenamefont
  {Campuzano}, \citenamefont {Norman}, \citenamefont {Ding}, \citenamefont
  {Randeria}, \citenamefont {Yokoya}, \citenamefont {Takeuchi}, \citenamefont
  {Takahashi}, \citenamefont {Mochiku}, \citenamefont {Kadowaki}, \citenamefont
  {Guptasarma},\ and\ \citenamefont {Hinks}}]{Campuzano98}%
  \BibitemOpen
  \bibfield  {author} {\bibinfo {author} {\bibfnamefont {J.~C.}\ \bibnamefont
  {Campuzano}}, \bibinfo {author} {\bibfnamefont {M.~R.}\ \bibnamefont
  {Norman}}, \bibinfo {author} {\bibfnamefont {H.}~\bibnamefont {Ding}},
  \bibinfo {author} {\bibfnamefont {M.}~\bibnamefont {Randeria}}, \bibinfo
  {author} {\bibfnamefont {T.}~\bibnamefont {Yokoya}}, \bibinfo {author}
  {\bibfnamefont {T.}~\bibnamefont {Takeuchi}}, \bibinfo {author}
  {\bibfnamefont {T.}~\bibnamefont {Takahashi}}, \bibinfo {author}
  {\bibfnamefont {T.}~\bibnamefont {Mochiku}}, \bibinfo {author} {\bibfnamefont
  {K.}~\bibnamefont {Kadowaki}}, \bibinfo {author} {\bibfnamefont
  {P.}~\bibnamefont {Guptasarma}}, \ and\ \bibinfo {author} {\bibfnamefont
  {D.~G.}\ \bibnamefont {Hinks}},\ }\href {\doibase 10.1038/32366} {\bibfield
  {journal} {\bibinfo  {journal} {Nature}\ }\textbf {\bibinfo {volume} {392}},\
  \bibinfo {pages} {157} (\bibinfo {year} {1998})}\BibitemShut {NoStop}%
\bibitem [{\citenamefont {Lee}(2014)}]{Lee14}%
  \BibitemOpen
  \bibfield  {author} {\bibinfo {author} {\bibfnamefont {P.~A.}\ \bibnamefont
  {Lee}},\ }\href {\doibase 10.1103/PhysRevX.4.031017} {\bibfield  {journal}
  {\bibinfo  {journal} {Phys. Rev. X}\ }\textbf {\bibinfo {volume} {4}},\
  \bibinfo {pages} {031017} (\bibinfo {year} {2014})}\BibitemShut {NoStop}%
\end{thebibliography}%

\end{document}


\setcounter{equation}{0}
\setcounter{figure}{0}
\setcounter{table}{0}
\setcounter{page}{1}
\makeatletter
\renewcommand{\theequation}{S\arabic{equation}}
\renewcommand{\thefigure}{S\arabic{figure}}

\title{Supplementary material for : Electronic spectral function in fractionalized Pair Density Wave scenario}

\author{M. Grandadam}

\affiliation{Institut de Physique Th\'eorique, Universit\'e Paris-Saclay, CEA, CNRS, F-91191 Gif-sur-Yvette, France}

\author{D. Chakraborty}

\affiliation{Institut de Physique Th\'eorique, Universit\'e Paris-Saclay, CEA, CNRS, F-91191 Gif-sur-Yvette, France}
\affiliation{Department of Physics and Astronomy, Uppsala University, Box 516, S-751 20 Uppsala, Sweden}\thanks{Present address}

\author{X. Montiel}

\affiliation{Department of Physics, Royal Holloway, University of London, Egham Surrey, United Kingdom}

\author{C. P\'epin}

\affiliation{Institut de Physique Th\'eorique, Universit\'e Paris-Saclay, CEA, CNRS, F-91191 Gif-sur-Yvette, France}

\maketitle

\section{Fluctuations of a PDW}\label{sec:Fluctuations-of-a}
The effective field theory for a fluctuating $\eta$-mode
is a rotor model with action\cite{Chakraborty19,Grandadam19} 
\begin{align}
S & =\frac{1}{2}\int d^{2}x\sum_{a,b=1}^{2}\left|\omega_{ab}\right|^2,\nonumber \\
\mbox{with } & \omega_{ab}=z_{a}^{*}\partial_{\mu}z_{b}-z_{b}\partial_{\mu}z_{a}^{*},\label{eq:rotor1}
\end{align}
with $z_{1}=\Delta_{ij}$, $z_{2}=\chi_{ij}$,
$z_{1}^{*}=\Delta_{ij}^{*}$, $z_{2}^{*}=\chi_{ij}^{*}$. The transformation
\begin{align}
z_{a} & \rightarrow z_{a}e^{i\theta},\label{eq:rotor2}
\end{align}
 leads to an additional term $2\partial_{\mu}\theta\left(\sum_{a}z_{a}^{*}z_{a}\right)$
into Eq.(\ref{eq:rotor1}). Imposing gauge invariance thus leads
to the constraint 
\begin{align}
\sum_{a}\left|z_{a}\right|^{2}=const,\label{eq:rotor3}
\end{align}
or equivalently 
\begin{align}
\left|\Delta_{ij}\right|^2+\left|\chi_{ij}\right|^2 & =const.\label{eq:rotor4}
\end{align}

\section{Derivation of the effective action for the doublet $\Psi_k$}
\label{sec:AppB}
We present here the derivation to obtain the effective action Eq.(5) starting for the real space Hamiltonian Eq.(4) of the main text. Our starting point is thus :
\begin{equation}
H = \sum_{i,j,\sigma} t_{ij} \left( c_{i \sigma}^{\dagger }c_{j \sigma}  + h.c \right) + J_{ij}\ \bm{S}_i \cdot \bm{S}_j + V_{ij}\ n_i n_j.   
\end{equation}
We start by going into momentum space by performing a Fourier transform of the electronic operator, the Hamiltonian becomes :
\begin{align}
H = &\sum_k \xi_k c^{\dagger}_{k \sigma}c_{k \sigma} +\sum_{k,k^{\prime},q} \sum_{\alpha,\mu} \left(V_q c^{\dagger}_{k, \alpha} c_{k+q, \alpha} c^{\dagger}_{k^{\prime}+q, \mu} c_{k^{\prime}, \mu}\right) \nonumber \\
&+\sum_{k,k^{\prime},q} \sum_{\alpha,\beta,\mu,\nu}\left( J_q \bm{\sigma}_{\alpha \beta} \cdot \bm{\sigma}_{\mu \nu} c^{\dagger}_{k, \alpha} c_{k+q, \beta} c^{\dagger}_{k^{\prime}+q, \mu} c_{k^{\prime}, \nu}\right),
\end{align}
where $\alpha,\beta,\mu$ and $\nu$ are spin indices and $\bm{\sigma}$ is the vector of Pauli matrices. We use $\xi_k = -2t\left(\cos\left(k_x\right)+\cos\left(k_y\right)\right) - 4 t^{\prime}\cos\left(k_x\right)\cos\left(k_y\right) - 2t^{\prime\prime}\left(\cos\left(2k_x\right)+\cos\left(2k_y\right)\right)-4t^{\prime\prime\prime}\left(\cos\left(2k_x\right)\cos\left(k_y\right)+\cos\left(2k_y\right)\cos\left(k_x\right)\right)-\mu$ with $t = 0.22$, $t^{\prime} = -0.034315$, $t^{\prime\prime} = 0.035977$, $t^{\prime\prime\prime} = -0.0071637$ and $\mu = -0.24327$ from Ref.\onlinecite{He11}. We can build from this Hamiltonian a path integral representation :
\begin{align}
S = \int & d\tau \sum_k  c^{\dagger}_k \left(i\partial_{\tau} - \xi_k \right)c_k \nonumber \\
& -\sum_{k,k^{\prime},q} \sum_{\alpha,\mu} \left(V_q c^{\dagger}_{k, \alpha} c_{k+q, \alpha} c^{\dagger}_{k^{\prime}+q, \mu} c_{k^{\prime}, \mu}\right) \nonumber \\ 
&-\sum_{k,k^{\prime},q} \sum_{\alpha,\beta,\mu,\nu} \left(J_q \bm{\sigma}_{\alpha \beta} \cdot \bm{\sigma}_{\mu \nu}  c^{\dagger}_{k, \alpha} c_{k+q, \beta} c^{\dagger}_{k^{\prime}+q, \mu} c_{k^{\prime}, \nu}\right).
\end{align}
We can now introduce two bosonic fields defined as $\Delta_{k} = \left\langle \sum_{\sigma} \sigma c_{k \sigma}c_{-k -\sigma}\right\rangle$ and $\chi_{k} = \left\langle \sum_{\sigma} c^{\dagger}_{k \sigma}c_{k+Q \sigma}\right\rangle$ we find the action: 
\begin{align}
\mathcal{S} &= \mathcal{S}_0 + \mathcal{S}_{\Delta} + \mathcal{S}_{\chi},\\
\mathcal{S}_0 &=  \int d \tau \sum_{k,\sigma} c^{\dagger}_{k \sigma} \left(i\partial_{\tau}-\xi_k \right) c_{k \sigma}, \\
\mathcal{S}_{\Delta} &=  \int d \tau \sum_{k,q}\left[\sum_{\sigma}\left( \sigma \Delta_{k+q} c^{\dagger}_{k \sigma} c^{\dagger}_{-k \bar{\sigma}} + \Delta^{*}_{k} c_{-k-q \bar{\sigma}} c_{k+q \sigma}\right)\right. \nonumber \\
& \quad + \left.\frac{\Delta_{k+q} \Delta_k^*}{J^-_q}\right], \\
\mathcal{S}_{\chi} &= \int d\tau \sum_{k,q}\left[\sum_{\sigma}\left( \chi_{k+q} c^{\dagger}_{k+Q \sigma} c_{k \sigma} + \chi^{*}_{k} c^{\dagger}_{k+q \sigma} c_{k+Q+q \sigma}\right)\right. \nonumber \\
& \quad + \left.\frac{\chi_{k+q} \chi_k^*}{J^+_q}\right],
\end{align}
Integrating the fermionic degree of freedom leads to an effective action for the bosonic fields:
\begin{align}
S = \int d \tau & \sum_{k,q}  \left(\frac{\Delta_{k+q} \Delta_k^*}{J^-_q} + \frac{\chi_{k+q} \chi_k^*}{J^+_q}\right) \nonumber \\
& - \text{Tr ln}\left( i\omega_n-\xi_k-\frac{|\Delta_k|^2}{i\omega_n+\xi_k} - \frac{|\chi_k|^2}{i\omega_n-\xi_{k+Q}}\right) \label{eq:Seff}
\end{align}
where $J_{\pm} = 3J \pm V$. We can now rewrite this effective action in term of the doublet field $\Psi_k = \left(\Delta_k,\ \chi_k\right)^T$ by using the identity $a |\Delta_k|^2 + b |\chi_k|^2 = \left(a+b\right)\frac{|\Psi_k|^2}{2} + \left(a-b\right) \frac{{\Psi_k}^{\dagger} \sigma^z \Psi_k}{2} $. We then use our ansatz for the pseudogap that the $|\Psi_k|$ acquires a finite amplitude but not $\Delta_k$ or $\chi_k$ and only keep the terms proportional to $|\Psi_k|^2$. This gives us the effective action Eq.(5) of the main text:
\begin{equation}
S_{eff} = \int d \tau \sum_{Q,k,q} \frac{{\Psi_{k}}^{\dagger} \Psi_{k+q}}{\tilde{J_q}} - \text{Tr}\ ln{\left(G^{-1}\left(i \omega, k\right)\right)}
\end{equation}
We can compare the result obtained by minimizing this action with respect to $|\Psi_k|$ to the more standard approach which consists in minimizing Eq.\eqref{eq:Seff} with respect to $|\Delta_k|$ or $|\chi_k|$. We report for this the condensation energy gained by the system close to the Fermi level:
\begin{align}
&E_{PG} = \frac{1}{2 \tilde{J}}|\Psi_{k=k_F}|^2 = 0.017\ eV, \\
&E_{SC} = \frac{1}{2 J_-}|\Delta_{k=k_F}|^2 = 0.014\ eV, \\
&E_{CDW} = \frac{1}{2 J_+}|\chi_{k=k_F}|^2 = 0.011\ eV.
\end{align}
The logic behind the result is that because of the density-density interaction V, the CDW order is prefered over the SC order close to the hot-spots. However this modulated order is limited to region in the Brillouin zone where the modulation wave-vector links different parts of the Fermi surface. The doublet is then able to have `the better of both world' as the coupling constant $\tilde{J}$ is between $J_+$ and $J_-$ but it is not as restricted as the CDW order and extends in a significant part of the anti-nodal region as shown in Fig.1 of the main text.
\section{Comparison with other scenario based on SC or CDW order}
We discuss here how the presence of a back-bending $k_G > k_F$ and the observation that the gap close from below when going towards the centre of the Brillouin zone are signatures for the presence of both particle-particle and modulated particle-hole pairs. In the following, we look at different origins for the gap in the ANR that lead only to one of the two features mentioned previously.

\subsubsection{Case of SC order}
\begin{figure}[t]
\includegraphics[width = 8.6 cm]{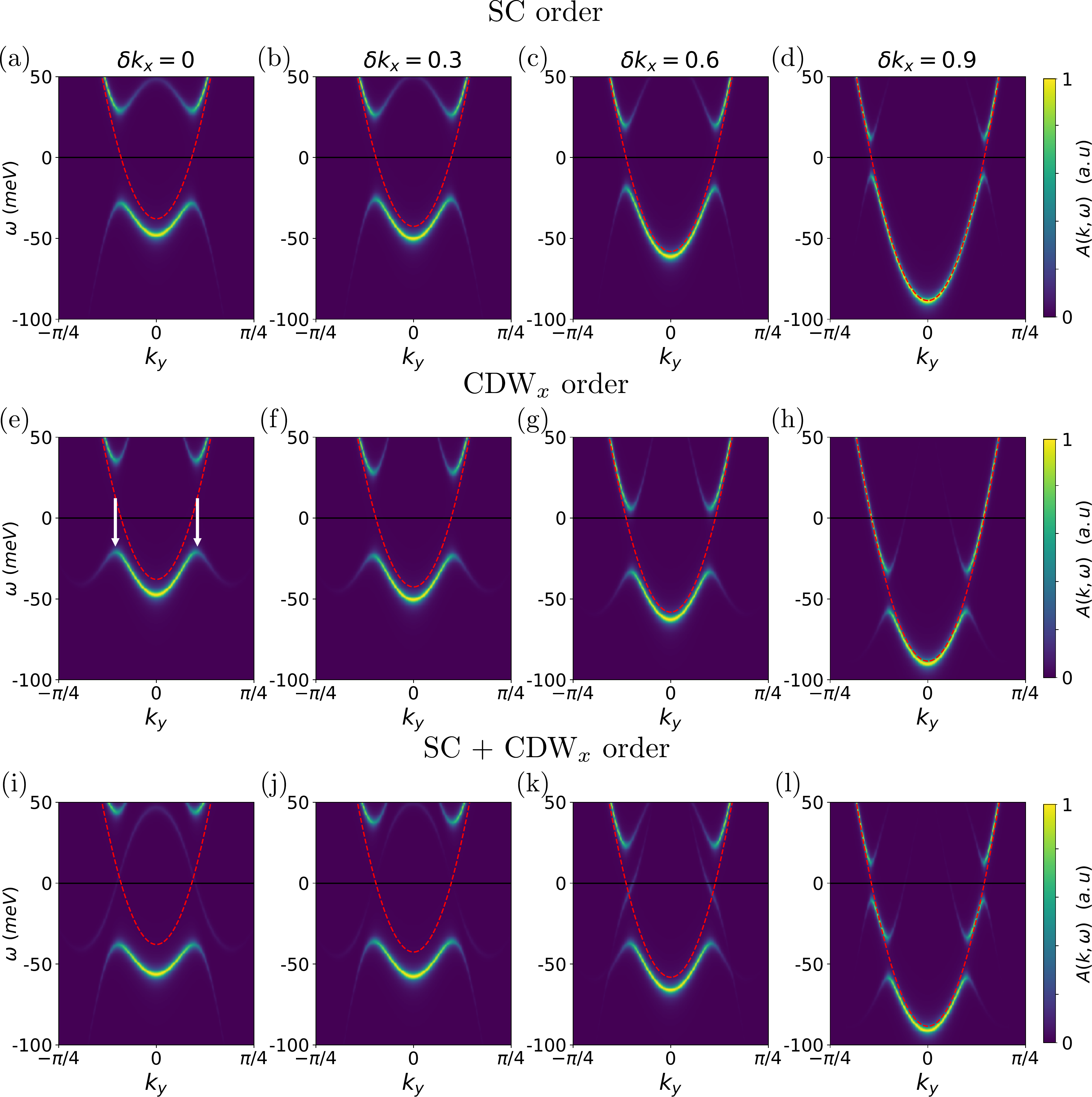}
\caption{\textbf{(a)-(d)} Energy dependence of the spectral function for cuts at fixed $k_x = \pi - \delta k_x$. This corresponds to the Green's function in Eqs.\eqref{eq:Gsc} and shows the effect of SC orer in the ANR. \textbf{(e)-(h)} Energy dependence of the spectral function for cuts at fixed $k_x = \pi - \delta k_x$. This corresponds to the Green's function in Eq.\eqref{eq:Gcdw} and shows the effect of CDW order with a wave-vector $Q = \left(\pm0.2,0\right)\pi$ in the ANR. \textbf{(i)-(l)} Energy dependence of the spectral function for cuts at fixed $k_x = \pi - \delta k_x$. This corresponds to the Green's function in Eq.\eqref{eq:Gsup} and shows the effect of the superposition of SC order and CDW order with a wave-vector $Q = \left(\pm0.2,0\right)\pi$ in the ANR. For all figures we used a gap of the form $|\Delta_k| = |\Delta_0|\ \text{exp}\left(-\frac{\left(k_x-\pi\right)^2}{2\sigma_x^2}-\frac{k_y^2}{2\sigma_y^2}\right)$ with $|\Delta_0| = 30\ meV$, $\sigma_x = 0.7$, $\sigma_y = 1$. The red lines indicate the non-interacting dispersion.}
\label{fig:A1}
\end{figure}

We start by reminding what is the consequence of a superconducting order on the electronic spectral function. For this, we consider a Green's function of the form 
\begin{equation}
    G_{SC}\left(k,i \omega_n \right) = \left(i\omega_n - \xi_k - \frac{|\Delta_k|^2}{i\omega_n+\xi_k}\right)^{-1}. \label{eq:Gsc}
\end{equation}
Fig.\ref{fig:A1}(a)-(d) shows the associated spectral function for different cut at fixed $k_x = \pi-\delta k_x$.
We see that the gap is closing from below as we go away from the edge of the Brillouin zone, however the back-bending of the band below the Fermi level is locked at $k_G = k_F$ due to particle-hole symmetry. This is in contrast with the experimental observation in Ref.\onlinecite{He11} and thus the pseudogap cannot be due solely to fluctuations of a superconducting order\cite{Norman:1998va,Chien2009,Campuzano98}.

\subsubsection{Case of CDW order}
It has been argued \cite{Lee14} that a pure CDW order cannot account for the experimental features observed in the ANR. We reproduce here the argument and show that a pure CDW order leads to a back-bending shifted from the original Fermi momentum but with a gap closing from above as we go to the centre of the Brillouin zone. We thus take a Green function of the form
\begin{equation}
    G_{CDW}\left(k,i \omega_n \right) = \left(i\omega_n - \xi_k - \frac{|\chi_k|^2}{i\omega_n-\xi_{k+Q}}\right)^{-1}. \label{eq:Gcdw}
\end{equation}
To be consistent with the experimental observations, we need to choose a modulation wave-vector along one of the cristal axis. As taking $Q$ along the $y$ axis will not lead to any gap in the ANR close to $k = \left(\pi,0\right)$ we take $Q = \left(\pm 0.2,0\right) \pi$. This effectively leads to a gap at the Fermi level as shown in Fig.\ref{fig:A1}(e)-(h) with a back-bending of the occupied band at $k_G > k_F$. However, we see that as we go closer to the center of the Brillouin zone, the gap is closing because the unoccupied band is going closer to the Fermi level. This result in a leftover gap for negative energy once $k_x > 0.2 \pi$.

\subsubsection{Coexistence of SC and CDW order}

\begin{figure}[t]
\center
\includegraphics[width = 8.6 cm]{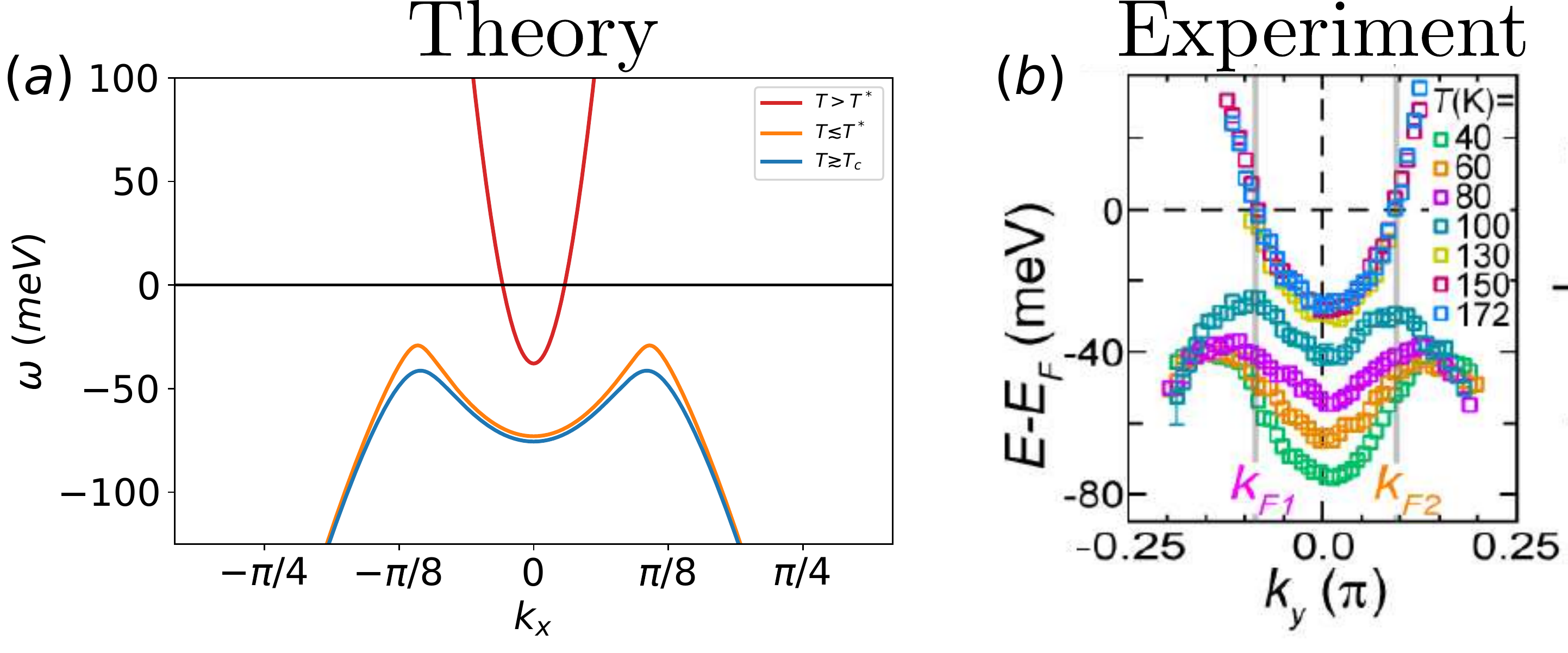}
\caption{(a) Temperature evolution of the band at the zone edge in the pseudogap regime. The red line indicates the non-interacting band above $T^*$. The orange line is the band after the opening of the pseudogap presenting a back-bending shifted from the original Fermi momentum $k_f$. When going down in temperature we add a finite mean-field amplitude for the SC order and obtain the band dispersion represented in blue. This leads to a change in the gap with respect to the Fermi level while the bottom of the band is unchanged. (b) Experimental measurement of the band at the zone edge for a range of temperature going from just above $T_c$ to above $T^*$.}
\label{fig:A2}
\end{figure}

Both of the previous examples do not recover the two main experimental features seen in ARPES in the ANR. Superconducting order gives a gap closing from below while a charge modulation leads to a shifted back-bending. We could hope that a coexistence of these order will bring the best of both worlds and allow for a description of the experiments. This is not the case however if we keep a modulation wave-vector along the $x$ axis as is shown in Fig.\ref{fig:A1}(i)-(l) where we considered a superposition of SC and CDW order along $x$. The Green's function in this case is the same as the one considered in the main text : 
\begin{align}
    G^{-1} \left( i \omega,k \right) &= i \omega -\xi_k - \sum_{Q = \pm Q_x, \pm Q_y}\frac{|\Psi_k|^2}{2} \tilde{G} \left( i\omega,k \right), \label{eq:Gsup} \\
    \tilde{G}\left(i \omega,k \right) &= \left(i \omega - \xi_{k+Q} \right)^{-1} + \left(i \omega + \xi_k \right)^{-1}.
\end{align}
 At the Brillouin zone edge, there is no crossing between the band from the superconducting order $-\xi_k$ and the band coming from the charge order $\xi_{k+Q}$. Because of this the band below the Fermi level has a back-bending at $k_G = k_F$ as in the superconducting case. The presence of charge order along the $x$ axis also leads to the presence of a gap below the Fermi level for $k_x > 0.2 \pi$. Moreover, we do not obtain the formation of a flat band at the bottom of the original dispersion, we will thus need another mechanism to explain the observation of this band below $T_c$. We can also note that all of these scenarii lead to a dispertion with a minimum close to the non-interacting band in contrast to what is observed experimentally (Fig.2 in the main text). In conclusion, despite not opening a gap by itself, choosing a modulation wave-vector along the $y$ direction coexisting with the SC order is the only possibility matching the experimental observation as shown in the main text (Fig.2).\\

\section{Effect of superconducting fluctuations}

In order to explain the temperature dependence of the dispersion between $T_c$ and $T^*$ we used the fact that CDW amplitude could acquire a long-range component thus modifying the electronic spectral function. Here we show that using the same argument with superconducting fluctuation leads to a very different result. The analogue of Fig.4 in the main text is shown in Fig.\ref{fig:A2}(a) where the red is the high-temperature dispersion ($T > T^*$) in the normal state, the orange line is the dispersion in the pseudogap for $T \lesssim T^*$ and the blue line is the dispersion for $T \gtrsim T_c$ if we consider long-range component for the superconducting order $|\Delta_k|$. We can see that a long-range component for the SC amplitude has an opposite effect than the long-range CDW amplitude. Namely, it induces a change in the gap by lowering the energy of the maximum of the band without affecting the energy of the bottom of the band. This is not in agreement with the experimental changes showed in Fig.\ref{fig:A2}(b)

\bibliographystyle{apsrev4-1}
\bibliography{Cuprates}